\documentclass[twocolumn,twocolappendix,times,resetfootnote]{aastex701}

\graphicspath{{./}{figures/}}

\usepackage{color} 
\usepackage{amsmath}
\usepackage{float}
\usepackage{amssymb}
\usepackage{graphicx}
\usepackage{natbib}
\usepackage{multirow}
\usepackage{array}
\usepackage{CJK} 

\newcommand{\blue}[1]{\textcolor{black}{#1}}

\definecolor{emerald}{rgb}{0.1,0.5,0.3}

\definecolor{purple}{HTML}{DA70D6}

\definecolor{gray}{HTML}{A9A9A9}

\newcommand{\squiggle}{SQuIGG$\vec{L}$E}
\newcommand{\mgii}{\ion{Mg}{2}}
\newcommand{\feii}{\ion{Fe}{2}}
\newcommand{\nad}{Na\,D}

\begin{document}
\begin{CJK*}{UTF8}{bsmi}
\title{\squiggle: Ubiquitous Cool Gas Outflows with Large Mass-Loading Factors in $z\sim0.7$ Post-starburst Galaxies}

\author[0000-0002-0696-6952]{Yuanze Luo}
\affiliation{Department of Physics and Astronomy and George P. and Cynthia Woods Mitchell Institute for Fundamental Physics and Astronomy, Texas A\&M University, 4242 TAMU, College Station, TX 77843-4242, US}
\email{yluo37@tamu.edu}

\author[0000-0003-3256-5615]{Justin S. Spilker}
\affiliation{Department of Physics and Astronomy and George P. and Cynthia Woods Mitchell Institute for Fundamental Physics and Astronomy, Texas A\&M University, 4242 TAMU, College Station, TX 77843-4242, US}
\email{jspilker@tamu.edu}

\author[0000-0003-4075-7393]{David J. Setton}\thanks{Brinson Prize Fellow}
\affiliation{Department of Astrophysical Sciences, Princeton University, Princeton, NJ 08544, USA}
\email{davidsetton@princeton.edu}

\author[0000-0001-6670-6370]{Timothy Heckman}
\affiliation{William H. Miller III Department of Physics and Astronomy, Johns Hopkins University, Baltimore, MD 21218, USA}
\affiliation{School of Earth and Space Exploration, Arizona State University, Tempe, AZ 85287, USA}
\email{theckma1@jhu.edu}


\author[0000-0001-5063-8254]{Rachel Bezanson}
\affiliation{Department of Physics and Astronomy and PITT PACC, University of Pittsburgh, Pittsburgh, PA 15260, USA}
\email{rachel.bezanson@pitt.edu}

\author[0000-0002-1759-6205]{Vincenzo R. D'Onofrio}
\affiliation{Department of Physics and Astronomy and George P. and Cynthia Woods Mitchell Institute for Fundamental Physics and Astronomy, Texas A\&M University, 4242 TAMU, College Station, TX 77843-4242, US}
\email{donofr19@tamu.edu}

\author[0009-0009-9810-3334]{Alex Geiger}
\affiliation{Department of Physics and Astronomy and George P. and Cynthia Woods Mitchell Institute for Fundamental Physics and Astronomy, Texas A\&M University, 4242 TAMU, College Station, TX 77843-4242, US}
\email{ageiger@tamu.edu}

\author[0000-0003-4700-663X]{Andy D. Goulding}
\affiliation{Department of Astrophysical Sciences, Princeton University, Princeton, NJ 08544, USA}
\email{goulding@astro.princeton.edu}

\author[0000-0002-5612-3427]{Jenny E. Greene}
\affiliation{Department of Astrophysical Sciences, Princeton University, Princeton, NJ 08544, USA}
\email{jennyg@princeton.edu}

\author[0000-0001-7883-8434]{Kate Rowlands}
\affiliation{AURA for ESA, Space Telescope Science Institute, 3700 San Martin Drive, Baltimore, MD 21218, USA}
\affiliation{William H. Miller III Department of Physics and Astronomy, Johns Hopkins University, Baltimore, MD 21218, USA}
\email{krowlands@stsci.edu}

\author[0000-0002-1714-1905]{Katherine A. Suess}
\affiliation{Department for Astrophysical \& Planetary Science, University of Colorado, Boulder, CO 80309, USA}
\email{suess@colorado.edu}

\author[0000-0001-8728-2984]{Elizabeth Taylor}
\affiliation{Institute for Astronomy, University of Edinburgh, Royal Observatory,Edinburgh EH9 3HJ, UK}
\email{etaylo2@roe.ac.uk}

\author[0000-0003-1535-4277]{Margaret E. Verrico}
\affiliation{University of Illinois Urbana-Champaign Department of Astronomy, University of Illinois, 1002 W. Green St., Urbana, IL 61801, USA}
\affiliation{Center for AstroPhysical Surveys, National Center for Supercomputing Applications, 1205 West Clark Street, Urbana, IL 61801, USA}
\email{verrico2@illinois.edu}

\author[0000-0002-1272-9064]{Andrew Weldon}
\affiliation{Department of Physics and Astronomy and George P. and Cynthia Woods Mitchell Institute for Fundamental Physics and Astronomy, Texas A\&M University, 4242 TAMU, College Station, TX 77843-4242, US}
\email{weldon@tamu.edu}  

\author[0000-0002-8956-7024]{Vivienne Wild}
\affiliation{School of Physics and Astronomy, University of St Andrews, North Haugh, St Andrews, KY16 9SS, UK}
\email{vw8@st-andrews.ac.uk}

\author[0000-0002-6768-8335]{Pengpei Zhu (朱芃佩)}
\affiliation{Cosmic Dawn Center (DAWN), Denmark}
\affiliation{DTU Space, Technical University of Denmark, Elektrovej 327, 2800 Kgs. Lyngby, Denmark}
\affiliation{INAF-Osservatorio Astrofisico di Arcetri, Largo Enrico Fermi 5, I-50125 Firenze, Italy}
\email{penzhu@space.dtu.dk}

\begin{abstract}



While recent James Webb Space Telescope studies have highlighted cool gas outflows as an important mechanism for rapid quenching at $z\gtrsim2$, such space-based spectroscopic samples remain limited in size. In this work, we search for and characterize cool gas outflows traced by \feii\ $\lambda\lambda$2586,2600 and \mgii\ $\lambda\lambda$2796,2803 absorption in post-starburst galaxies (PSBs) at $0.5<z<0.9$, which are accessible to ground-based spectroscopy. Utilizing the \squiggle\ survey containing over 1300 PSBs, we stack spectra in bins of stellar mass, (specific) star formation rate, burst mass, burst mass fraction, and time since quenching. We observe blueshifted line centroids indicative of outflows in all stacks except the one containing galaxies that quenched more than 260 Myr ago. Outflow velocities decline with increasing time since quenching, while showing little correlation with other galaxy properties. The inferred outflows have large mass loading factors of \blue{1 -- 120}, maximum velocities above the interstellar medium escape velocities of their host galaxies, and energy and momentum fluxes exceeding those expected from the current star formation. Our results suggest that cool gas outflows are ubiquitous in intermediate-redshift PSBs and are likely relic outflows launched by previous starburst or AGN activity. These outflows have the potential to quench star formation by transporting gas into the galactic halos, consistent with outflows playing an important role in rapid quenching from intermediate to high redshift.


\end{abstract}



\section{Introduction}

Observations of galaxies have revealed a well-established bimodality in both morphology and color up to $z\sim3$ \citep[e.g.,][]{strat2001,Baldry_2004,Willmer_2006,Brammer_2009,van_der_Wel_2014}: the majority of galaxies are either in the ``blue cloud" (spiral, star-forming) or the ``red sequence'' (elliptical, quiescent). Galaxies appear to evolve from blue to red \citep[e.g.,][]{Bell_2004,Faber_2007,Weaver_2023} as star formation (SF) shuts down, a process referred to as ``quenching" (see \citealt{Whitaker_2026} for a review). The structural transformation and cessation of SF appear closely linked, and galaxies need to eventually lose their gas reservoirs to become red and dead.

Galactic-scale outflows have been invoked in theoretical models and simulations of galaxy evolution to explain observed properties of massive galaxies including the chemical enrichment of the circumgalactic medium (CGM) and the existence of red, gas-deprived galaxies \citep[e.g.,][]{Silk_1998, Thompson_2005, Oppenheimer_2010, Nelson_2019}. These outflows regulate the baryon cycle by injecting energy and momentum into gas reservoirs of galaxies, and have been observed in galaxies across a wide range of evolutionary stages in different phases \citep[e.g.,][]{Heckman_2000, Rupke_2005a, Harrison_2014, Fluetsch_2019,Thompson_2024}.

While outflows appear crucial in shutting down SF, especially in the cases of rapid quenching \citep[e.g.,][]{Springel_2005,Spilker_2022}, the direct causal link between outflows and quenching and the degree of impact of outflows remain uncertain. An effective way to investigate this question is to study outflows in post-starburst galaxies (PSBs), systems that have undergone rapid and recent quenching within the past 1 Gyr. PSBs are ideal laboratories for studying quenching mechanisms and are typically identified by strong Balmer absorption (signature of large populations of intermediate-age stars) and lack of emission lines (indication of low current SF) in their spectra (see \citealt{French_2021} for a review). Restframe UV-optical absorption lines (e.g., from Fe, Mg, Na) are widely used tracers for cool gas outflows and have been studied in PSBs at both low redshift ($z\lesssim0.3$; e.g., \citealt{Baron_2020,Baron_2022,luo2022,Sun_2024}) and high redshift ($z\gtrsim1.5$; e.g., \citealt{Maltby_2019,Taylor_2024,Davies_2024,Wu_2025,Valentino_2025,Taylor_2026}). These studies show mixed results on the impact of outflows, with some reporting high-velocity winds capable of escaping their host galaxies and others finding more modest velocities. 

Recent work on rapidly quenched galaxies beyond cosmic noon has gone further by estimating outflow masses, showing that the outflows, though not fast enough to expel materials from the host galaxies, have mass outflow rates an order of magnitude larger than the star formation rates (SFR). This result suggests that such outflows could efficiently transport gas into the CGM, consistent with a ``galactic fountain'' scenario, and may represent a dominant mechanism for the rapid quenching of massive galaxies at $z \gtrsim 2$ \citep{Davies_2024,Belli_2024,Wu_2025,Zhu_2026}. These studies speculate that these outflows are relics of past active galactic nuclei (AGN) activity, further suggesting a connection between AGN feedback and rapid quenching.

Although studies at $z \gtrsim 2$ provide intriguing evidence, they mostly rely on expensive space-based spectroscopy and are limited in sample statistics. In this work, we focus on the intermediate redshift regime, $0.5<z<0.9$, where large ground-based spectroscopic surveys such as the Sloan Digital Sky Survey (SDSS) enable the selection of over a thousand PSBs while covering the \feii\ $\lambda\lambda$2586,2600, \mgii\ $\lambda\lambda$2796,2803, and \nad\ $\lambda\lambda$5889,5895 absorption lines which trace cool (T $\sim10^4$ K), neutral and partially ionized gas. Our study will assess how common cool gas outflows are throughout the quenching process and explore their correlations with galaxy properties, complementing the limited sample size available at higher redshift and providing an important bridge to studies in the local universe.

This paper is organized as follows. In Section \ref{sec: sample} we outline the PSB sample and data used in this study. In Section \ref{sec: analysis} we describe our analysis approach in spectral stacking (Section \ref{sec: stack}), line profile modeling (Section \ref{sec: line model}), derivation of outflow properties (Section \ref{sec: outflow derivations}), and associated uncertainties (Section \ref{sec: outflow uncertainties}). We present our results on the comparison between outflow and galaxy properties in Section \ref{sec: outflow properties}, and further estimate the outflow energetics in Section \ref{sec: outflow energetics}. In Section \ref{sec: discussion}, we discuss possible outflow driving mechanisms, fate of the outflows, comparison with literature studies, and implications for quenching. We summarize our findings in Section \ref{sec: summary}. Throughout this paper, we assume a flat cosmological model with $H_0=70\ \rm{km\ s^{-1}\ Mpc^{-1}}$, $\Omega_m=0.3$, and $\Omega_{\Lambda}=0.7$. 

\section{Sample and Data} \label{sec: sample}

This work focuses on the \squiggle\ (Studying Quenching in Intermediate-z Galaxies: Gas, angu$\vec{L}$ar momentum, and Evolution) survey, which consists of more than 1318 PSBs selected from the SDSS DR14 spectroscopic sample \citep{Abolfathi_2018}. The sample selection and galaxy properties are described in detail in \citet{Suess_2022}. Here we provide a brief summary and refer readers to their work for more information.

The \squiggle\ survey selects galaxies at intermediate redshift ($0.5<z<0.9$) that have recently quenched their primary epoch of SF using two restframe color cuts. The colors are measured using the restframe $U_m$, $B_m$, and $V_m$ medium-band filters from \citet{Kriek_2010} and the cuts are designed to select galaxies with strong Balmer breaks (indicated by red $U_m-B_m$ colors) and blue slopes redward of the break (indicated by blue $B_m-V_m$ colors). \citet{Suess_2022} performed spectro-photometric fitting to infer the stellar population properties and nonparametric star formation histories (SFHs) of all galaxies in the sample. The fitting provides galaxy physical properties which we later use to decide stacking bins (Section \ref{sec: analysis}). The \squiggle\ PSBs turn out to be very massive (stellar mass $M_* > 10^{11} M_{\odot}$) and rapidly quenched, with small current star formation rates (SFR $\lesssim1 M_{\odot}$ yr$^{-1}$). The resulting SFHs confirm that the above color cuts selects PSBs that quenched within the last $\sim500$ Myr with 75\% of them having formed $>$25\% of their stellar mass in a recent burst (see \citealt{Suess_2022} for more details).

We utilize SDSS DR14 spectra for the spectral analysis, which covers \feii\ and \mgii\ for all \squiggle\ PSBs, and \nad\ for $\sim80$\% of the sample. The individual spectra are noisy at the absorption line regions. Using the inverse variance arrays provided with the spectra, we estimate that the mean signal noise ratio (SNR) at the absorption line regions are $\lesssim2$ for the majority of the sample, with a median $\sim1.5$. To further assess the strength of the absorption feature, we define SNR$_{\rm abs}$, which is calculated as the maximum absorption depth (deepest point below the continuum level) over the standard deviation of feature-free regions surrounding the line\footnote{Line regions for \feii,\, \mgii,\, and \nad\ are 2577--2606\AA\, 2781--2818\AA, and 5877--5912\AA. Feature-free regions for noise estimation are 2540--25\blue{60}\AA\ and 2635--2660\AA\ for \feii\ (avoiding nearby Mn \blue{and \feii*} lines), 2740--2770\AA\ and 2825--2860\AA\ for \mgii, 5800--5850\AA\ and 5920--5950\AA\ for \nad.} in the continuum-normalized spectra. Approximating the local continuum level as a straight line, we find that SNR$_{\rm abs}$ are $\lesssim3$ for the majority of the sample, with a median $\sim2.2$. We therefore opt for a stacking approach to create composite spectra with higher SNR for more detailed analysis, as described in the following section.



\section{Analysis} \label{sec: analysis}


To achieve high SNR at the absorption features and explore possible correlations between outflow and host galaxy properties, we make stacked spectra in different bins of galaxy properties. The galaxy properties we stack in are time since quenching ($t_{q}$), burst mass ($M_{\mathrm{burst}}$), burst mass fraction ($f_{\mathrm{burst}}$), $M_{*}$, SFR, and specific SFR (sSFR = SFR/$M_{*}$), which are from the catalog in \citet{Suess_2022}. $t_q$ tells how long the galaxy has been quenched and is defined as the time when the recent starburst ended determined by when the time derivative of the SFH drops below a threshold. $M_{\mathrm{burst}}$ and $f_{\mathrm{burst}}$ are the stellar mass formed in the burst and the fraction of the total stellar mass formed during the burst. We refer the readers to \citet{Suess_2022} and \citet{Suess_2022_sfh_model} for details about SFH models and spectro-photometric fitting.

To ensure similar SNR in the stacks, we divide the whole sample in 4 bins based on the 25/50/75th percentiles of $t_q$, $M_{\mathrm{burst}}$, $M_*$, and $f_{\mathrm{burst}}$ such that each stack contains about the same number of spectra. SFRs of the \squiggle\ sample span a narrow range due to the quenched nature of the galaxies, and \citet{Suess_2022} reported that SFRs below 1 M$_{\odot}$ yr$^{-1}$ should be treated as upper limits. Thus we stack only in 2 bins for SFR and sSFR, \blue{separated by SFR smaller than 1 and greater than 10 M$_{\odot}$ yr$^{-1}$, to sample the lower and higher ends of the distribution and test for potential trends between SFR and outflow properties. The SFR stack contains $\sim700$ (100) galaxies in the lower (higher) SFR bin, and stacks divided by the 25/50/75th percentiles of the corresponding quantities typically contain $\sim300$ galaxies each.} Information about the stacking bins are summarized in Table \ref{tab:outflow_properties} and we present our stacking and modeling approaches in the following sections.

\subsection{Spectra Stacking} \label{sec: stack}

\begin{figure*}
\centering
\includegraphics[width=\textwidth]{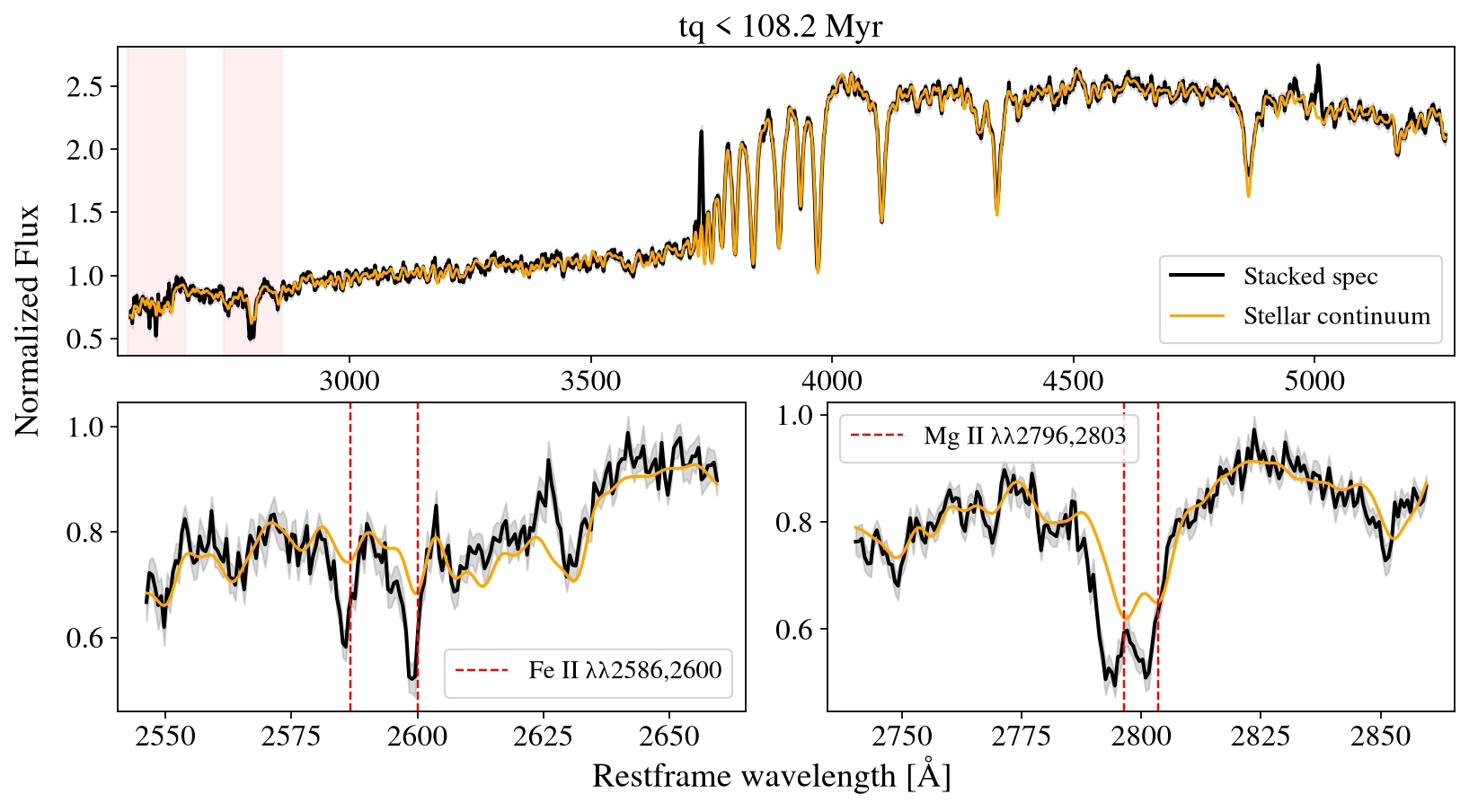}
\caption{Example of one stacked spectrum (the smallest t$_q$ group) and corresponding stellar continuum. The top panel shows the entire wavelength range while the bottom panels show zoomed-in views of the \feii\ and \mgii\ regions (also shaded in pink in the top panel). The uncertainty of the stacked spectrum is plotted as gray shaded regions, and the restframe wavelengths of \feii\ and \mgii\ lines are indicated by red vertical dashed lines. 
\label{fig:ppxf_example}}
\end{figure*}

Since all spectra cover both the \feii\ and \mgii\ absorption lines, we include the entire sample when constructing the \feii+\mgii\ stacks used in the main analysis. For \nad\ stacks, we follow the same stacking procedure detailed below, except removing galaxies ($\sim$20\% of the sample) whose spectra does not cover the \nad\ absorption lines.

To create the stack, we first shift each spectrum to the restframe and then interpolate it to a common wavelength grid covered by all spectra in the stack. \blue{We choose a wavelength spacing ($\Delta$log$\lambda$[\AA] = 10$^{-4}$) same as} that of the original SDSS spectra to avoid down-sampling during the stacking. The spectra are then normalized by their mean fluxes at the flat region of \blue{3000--3100\AA\ (for \feii+\mgii\ stacks) or 4150--4250\AA\ (for \nad\ stacks)} before final combination by taking the median flux at each wavelength. To avoid outliers affecting the final stacked spectrum, we undertake a sigma-clipping procedure following \citet{Taylor_2024}, removing input spectra that have more than 10\% flux points between 2540--2650\AA\ (the \feii\ region) and 2750--2850\AA\ (the \mgii\ region) lying at least 3$\sigma$ away from the initial \blue{\feii+\mgii} median stack \blue{(5800--5950\AA\ for the \nad\ stacks)}\footnote{This procedure removes less than 1\% of spectra in each stacking bin.}. The remaining spectra are then combined to create the final median stack. The flux uncertainty in the final median stack is estimated via bootstrapping: we make another 200 median stacks by drawing randomly with replacement from the input spectra the same number of spectra as in the final median stack, and take the standard deviation of the 200 stacked spectra as the 1$\sigma$ error. The mean SNR at the absorption line regions in the stacked spectra are $\gtrsim20$, much improved compared to the original individual spectra.

\begin{deluxetable*}{lcccccc}
\tablecaption{Galaxy stack properties and derived outflow quantities. \label{tab:outflow_properties}}
\tablehead{
\colhead{\blue{Stack Property}} &
\colhead{\blue{$v_{\rm out}$ [km\,s$^{-1}$]}} &
\colhead{\blue{$b_{\rm D}$ [km\,s$^{-1}$]}} &
\colhead{\blue{log($N_{\rm H}$ [cm$^{-2}$])}} &
\colhead{\blue{$C_f$}} &
\colhead{\blue{log($M_{\rm out}$ [$M_{\odot}$])}} &
\colhead{\blue{$\dot{M}_{\rm out}$ [$M_{\odot}\,{\rm yr}^{-1}$]}}
}
\startdata
\blue{log($M_*$ [$M_{\odot}$])} & & & & & & \\
\blue{$\leq 11.1$} & \blue{$-143^{+14}_{-13}$, $-277^{+29}_{-27}$} & \blue{$109^{+11}_{-6}$, $399^{+54}_{-44}$} & \blue{$21.3^{+0.2}_{-0.2}$, $19.6^{+0.2}_{-0.2}$} & \blue{$0.2^{+0.01}_{-0.01}$, $0.4^{+0.17}_{-0.07}$} & \blue{$8.9^{+0.12}_{-0.12}$, $7.4^{+0.07}_{-0.07}$} & \blue{$31.3^{+8.8}_{-8.8}$, $2.3^{+0.5}_{-0.5}$} \\
\blue{$(11.1, 11.2]$} & \blue{$-203^{+20}_{-23}$, $-327^{+37}_{-28}$} & \blue{$162^{+46}_{-29}$, $337^{+86}_{-49}$} & \blue{$21.0^{+0.2}_{-0.3}$, $19.6^{+0.2}_{-0.3}$} & \blue{$0.2^{+0.06}_{-0.01}$, $0.3^{+0.20}_{-0.06}$} & \blue{$8.5^{+0.16}_{-0.16}$, $7.4^{+0.09}_{-0.09}$} & \blue{$20.4^{+7.5}_{-7.5}$, $2.4^{+0.7}_{-0.7}$} \\
\blue{$(11.2, 11.3]$} & \blue{$-110^{+17}_{-18}$, $-370^{+15}_{-15}$} & \blue{$180^{+22}_{-18}$, $260^{+19}_{-18}$} & \blue{$21.2^{+0.1}_{-0.1}$, $19.7^{+0.1}_{-0.1}$} & \blue{$0.2^{+0.02}_{-0.01}$, $0.3^{+0.04}_{-0.03}$} & \blue{$8.7^{+0.09}_{-0.09}$, $7.5^{+0.06}_{-0.06}$} & \blue{$17.6^{+4.0}_{-4.0}$, $3.1^{+0.5}_{-0.5}$} \\
\blue{$>11.3$} & \blue{$-191^{+51}_{-63}$, $-387^{+21}_{-21}$} & \blue{$293^{+62}_{-54}$, $305^{+37}_{-27}$} & \blue{$21.2^{+0.2}_{-0.2}$, $20.0^{+0.1}_{-0.2}$} & \blue{$0.2^{+0.05}_{-0.03}$, $0.2^{+0.03}_{-0.02}$} & \blue{$8.7^{+0.13}_{-0.13}$, $7.6^{+0.12}_{-0.12}$} & \blue{$27.9^{+10.0}_{-10.0}$, $4.4^{+1.2}_{-1.2}$} \\
\hline
\blue{log(SFR [$M_{\odot}$ yr$^{-1}$])} & & & & & & \\
\blue{$<0$} & \blue{$-162^{+13}_{-13}$, $-168^{+22}_{-22}$} & \blue{$121^{+17}_{-13}$, $557^{+62}_{-63}$} & \blue{$21.3^{+0.1}_{-0.1}$, $19.5^{+0.2}_{-0.2}$} & \blue{$0.2^{+0.01}_{-0.01}$, $0.5^{+0.26}_{-0.12}$} & \blue{$8.8^{+0.10}_{-0.10}$, $7.5^{+0.05}_{-0.05}$} & \blue{$31.4^{+7.9}_{-7.9}$, $1.5^{+0.3}_{-0.3}$} \\
\blue{$>1$} & \blue{$-82^{+29}_{-28}$, $-448^{+51}_{-54}$} & \blue{$206^{+38}_{-36}$, $357^{+66}_{-57}$} & \blue{$21.1^{+0.4}_{-0.4}$, $19.0^{+0.5}_{-0.2}$} & \blue{$0.3^{+0.15}_{-0.06}$, $0.5^{+0.32}_{-0.30}$} & \blue{$8.8^{+0.15}_{-0.15}$, $6.9^{+0.06}_{-0.06}$} & \blue{$13.9^{+7.2}_{-7.2}$, $1.2^{+0.2}_{-0.2}$} \\
\hline
\blue{log(sSFR [yr$^{-1}$])} & & & & & & \\
\blue{$<-10.9$} & \blue{-\tablenotemark{a}} & \blue{-} & \blue{-} & \blue{-} & \blue{-} & \blue{-} \\
\blue{$>-10.5$} & \blue{-} & \blue{-} & \blue{-} & \blue{-} & \blue{-} & \blue{-} \\
\hline
\blue{$t_q$ [Myr]} & & & & & & \\
\blue{$\leq$ 108.2} & \blue{$-172^{+17}_{-21}$, $-404^{+16}_{-17}$} & \blue{$160^{+29}_{-21}$, $305^{+22}_{-22}$} & \blue{$21.2^{+0.2}_{-0.1}$, $19.4^{+0.2}_{-0.3}$} & \blue{$0.2^{+0.03}_{-0.03}$, $0.4^{+0.21}_{-0.08}$} & \blue{$8.7^{+0.09}_{-0.09}$, $7.2^{+0.07}_{-0.07}$} & \blue{$28.6^{+6.7}_{-6.7}$, $1.9^{+0.3}_{-0.3}$} \\
\blue{$(108.2, 169.0]$} & \blue{$-149^{+20}_{-22}$, $-385^{+13}_{-14}$} & \blue{$176^{+41}_{-28}$, $273^{+20}_{-17}$} & \blue{$21.2^{+0.1}_{-0.1}$, $19.7^{+0.1}_{-0.1}$} & \blue{$0.2^{+0.01}_{-0.01}$, $0.3^{+0.04}_{-0.03}$} & \blue{$8.7^{+0.11}_{-0.11}$, $7.5^{+0.06}_{-0.06}$} & \blue{$22.0^{+5.8}_{-5.8}$, $3.3^{+0.5}_{-0.5}$} \\
\blue{$(169.0, 266.0]$} & \blue{$-122^{+31}_{-34}$, $-246^{+36}_{-32}$} & \blue{$297^{+54}_{-44}$, $428^{+60}_{-56}$} & \blue{$21.1^{+0.1}_{-0.2}$, $19.4^{+0.2}_{-0.3}$} & \blue{$0.2^{+0.05}_{-0.01}$, $0.5^{+0.29}_{-0.14}$} & \blue{$8.6^{+0.10}_{-0.10}$, $7.4^{+0.06}_{-0.06}$} & \blue{$16.0^{+5.2}_{-5.2}$, $1.7^{+0.3}_{-0.3}$} \\
\blue{$>$ 266.0} & \blue{\nodata\tablenotemark{b}, $-36^{+41}_{-41}$} & \blue{\nodata, $963^{+65}_{-63}$} & \blue{\nodata, $19.4^{+0.1}_{-0.1}$} & \blue{\nodata, $0.8^{+0.14}_{-0.18}$} & \blue{\nodata, $7.5^{+0.02}_{-0.02}$} & \blue{\nodata, $0.3^{+0.3}_{-0.3}$} \\
\hline
\blue{$M_{burst}$ [$M_{\odot}$]} & & & & & & \\
\blue{$\leq 10.7$} & \blue{$-144^{+18}_{-18}$, $-355^{+19}_{-19}$} & \blue{$170^{+22}_{-19}$, $297^{+30}_{-28}$} & \blue{$21.3^{+0.1}_{-0.1}$, $19.5^{+0.2}_{-0.3}$} & \blue{$0.2^{+0.01}_{-0.01}$, $0.3^{+0.18}_{-0.07}$} & \blue{$8.8^{+0.09}_{-0.09}$, $7.2^{+0.09}_{-0.09}$} & \blue{$28.6^{+6.4}_{-6.4}$, $1.8^{+0.4}_{-0.4}$} \\
\blue{$(10.7, 10.9]$} & \blue{$-190^{+23}_{-41}$, $-353^{+27}_{-27}$} & \blue{$142^{+38}_{-28}$, $361^{+46}_{-38}$} & \blue{$21.4^{+0.5}_{-0.3}$, $19.6^{+0.2}_{-0.3}$} & \blue{$0.2^{+0.04}_{-0.03}$, $0.3^{+0.17}_{-0.06}$} & \blue{$8.9^{+0.20}_{-0.20}$, $7.3^{+0.09}_{-0.09}$} & \blue{$41.7^{+19.7}_{-19.7}$, $2.3^{+0.5}_{-0.5}$} \\
\blue{$(10.9, 11.0]$} & \blue{$-102^{+26}_{-31}$, $-324^{+29}_{-25}$} & \blue{$228^{+42}_{-32}$, $371^{+60}_{-44}$} & \blue{$21.1^{+0.1}_{-0.1}$, $19.8^{+0.1}_{-0.2}$} & \blue{$0.2^{+0.02}_{-0.01}$, $0.3^{+0.08}_{-0.03}$} & \blue{$8.7^{+0.09}_{-0.09}$, $7.5^{+0.10}_{-0.10}$} & \blue{$15.7^{+4.3}_{-4.3}$, $3.3^{+0.9}_{-0.9}$} \\
\blue{$>11.0$} & \blue{$-194^{+37}_{-37}$, $-360^{+27}_{-23}$} & \blue{$266^{+71}_{-62}$, $320^{+48}_{-35}$} & \blue{$21.3^{+0.3}_{-0.3}$, $19.9^{+0.1}_{-0.2}$} & \blue{$0.2^{+0.05}_{-0.03}$, $0.3^{+0.05}_{-0.03}$} & \blue{$8.7^{+0.14}_{-0.14}$, $7.6^{+0.10}_{-0.10}$} & \blue{$29.8^{+11.2}_{-11.2}$, $3.9^{+1.0}_{-1.0}$} \\
\hline
\blue{$f_{burst}$} & & & & & & \\
\blue{$\leq0.2$} & \blue{$-139^{+20}_{-23}$, $-361^{+18}_{-18}$} & \blue{$207^{+30}_{-26}$, $303^{+32}_{-28}$} & \blue{$21.2^{+0.2}_{-0.2}$, $19.8^{+0.2}_{-0.2}$} & \blue{$0.2^{+0.04}_{-0.03}$, $0.3^{+0.05}_{-0.03}$} & \blue{$8.8^{+0.11}_{-0.11}$, $7.4^{+0.09}_{-0.09}$} & \blue{$23.2^{+6.0}_{-6.0}$, $3.0^{+0.6}_{-0.6}$} \\
\blue{$(0.2, 0.3]$} & \blue{$-114^{+28}_{-29}$, $-358^{+19}_{-19}$} & \blue{$207^{+36}_{-32}$, $285^{+29}_{-26}$} & \blue{$21.2^{+0.3}_{-0.3}$, $19.7^{+0.2}_{-0.2}$} & \blue{$0.2^{+0.06}_{-0.04}$, $0.3^{+0.07}_{-0.04}$} & \blue{$8.7^{+0.14}_{-0.14}$, $7.4^{+0.09}_{-0.09}$} & \blue{$17.9^{+5.5}_{-5.5}$, $2.7^{+0.6}_{-0.6}$} \\
\blue{$(0.3, 0.5]$} & \blue{$-198^{+18}_{-17}$, $-370^{+27}_{-25}$} & \blue{$120^{+22}_{-14}$, $379^{+68}_{-55}$} & \blue{$21.4^{+0.2}_{-0.2}$, $19.6^{+0.2}_{-0.3}$} & \blue{$0.2^{+0.01}_{-0.00}$, $0.4^{+0.22}_{-0.07}$} & \blue{$8.9^{+0.15}_{-0.15}$, $7.4^{+0.09}_{-0.09}$} & \blue{$50.0^{+17.9}_{-17.9}$, $2.8^{+0.7}_{-0.7}$} \\
\blue{$>0.5$} & \blue{$-129^{+47}_{-57}$, $-306^{+33}_{-31}$} & \blue{$228^{+74}_{-57}$, $396^{+66}_{-49}$} & \blue{$21.3^{+0.6}_{-0.6}$, $19.8^{+0.1}_{-0.2}$} & \blue{$0.1^{+0.14}_{-0.04}$, $0.3^{+0.08}_{-0.03}$} & \blue{$8.6^{+0.22}_{-0.22}$, $7.5^{+0.09}_{-0.09}$} & \blue{$16.6^{+10.7}_{-10.7}$, $3.2^{+0.8}_{-0.8}$} \\
\enddata
\tablecomments{\blue{In each column of the outflow property, the first number is the measurement from \feii\ and the second number is that from \mgii. Values for column density and mass should be regarded as lower limits (see Section \ref{sec: outflow uncertainties} for more details).
Columns are as follows:
(1) galaxy properties in each stack;
(2) outflow velocity defined as the absorption-line centroid offset ($\Delta v$);
(3) Doppler line width $b_{\rm D} = \sqrt{2}\sigma$;
(4) hydrogen column density of the outflowing gas;
(5) covering fraction; 
(6) inferred outflow mass; and 
(7) mass outflow rate. 
\tablenotetext{a}{The sSFR stacks are the same as the SFR stacks and thus have the same outflow measurements.}
\tablenotetext{b}{Insufficient \feii\ SNR for fitting.}}}
\end{deluxetable*}

\subsection{Absorption Line Profile Modeling}\label{sec: line model}

To model the absorption line profiles originating from the interstellar medium (ISM), we first normalize the stacked spectra with their stellar continua. We fit the stellar continuum of each stacked spectrum with \texttt{pPXF} \citep{Cappellari_2004,Cappellari_2017}, using the \blue{GALAXEV} stellar population models\footnote{We also model the stellar continuum using the E-MILES \citep{Vazdekis_2016} and FSPS \citep{Conroy_2009} models, which produce fits comparable to those from \blue{GALAXEV}. We adopt the \blue{GALAXEV} stellar continua because they provide the best fit around the \mgii\ absorption feature and neighboring line-free regions.} \citep{Bruzual_2003}. We mask major emission and absorption features (e.g., H$\beta$, [\ion{O}{2}]$\lambda\lambda$3727,3729, [\ion{O}{3}]$\lambda\lambda$4959,5007), including in the mask \mgii, \feii, and \nad\ regions when fitting the stellar continuum, and include additive and multiplicative Legendre polynomials of degree 10 to account for the mismatch in shape between template and observed spectra as well as systematics in the flux calibration. An example of a stacked spectrum with its corresponding stellar continuum are shown in Figure \ref{fig:ppxf_example}. We verify the quality of the stellar continuum fitting by measuring the equivalent width (EW) of the Mg\,b absorption lines, which are dominated by stellar photospheric absorption \citep[e.g.,][]{Deeming_1960}, in the continuum-normalized spectra. The residual EW$_{\rm Mgb}$ measured within a $\pm15$\AA\ window encompassing the lines are $\lesssim0.2$\AA, in most cases $\lesssim0.1$\AA, indicating good continuum fitting.

Before modeling the absorption lines, we check that there is sufficient \blue{gas} absorption left after removing the stellar contributions. We measure both the EW within a $\pm15$\AA\ window encompassing lines and calculate the SNR$_{\rm abs}$ as defined in Section \ref{sec: sample}. We proceed to fitting if EW $>1$\AA\ and SNR$_{\rm abs}>4$. \mgii\ in all stacked spectra, and \feii\ in all but the largest $t_q$ stack satisfy these criteria, with large EW $\gtrsim2$\AA. In contrast, there is little \nad\ absorption from the ISM, and more than 80\% of the stacked spectra fail the quality check. We explore the implications of the non-detection of ISM \nad\ for the outflow mass limit in Appendix \ref{appendix mout}, and focus hereafter on the analysis of \mgii\ and \feii.

We model the \feii\ and \mgii\ absorption lines with Voigt profiles following \citet{Rupke_2005a}. To account for possible emission, we model the emission component as an additional Gaussian \cite[e.g.,][]{Tremonti_2007,Shaban_2025}. The fitting is performed utilizing a Bayesian inference method with the \texttt{emcee}\footnote{\url{https://emcee.readthedocs.io/en/stable/}} package \citep{mcmc}. We briefly outline the model below and refer the readers to corresponding works (and references therein) for more details.

The absorption profile in the continuum-normalized spectrum is defined as 
\begin{equation}
    I(\lambda) = 1 - C_f + C_f \times e^{-\tau_{\rm{B}}(\lambda)-\tau_{\rm{R}}(\lambda)},
\end{equation}
where $C_f$ is the velocity-independent covering fraction, related to the clumpiness of the gas along the line of sight, and $\tau_{\rm{B}}(\lambda)$ and $\tau_{\rm{R}}(\lambda)$ are the optical depths of the \mgii\ (\feii) doublet at 2796.3543 (2586.6500) and 2803.5315 (2600.1729) \AA, respectively. The optical depth is written as 
\begin{equation}
    \tau(\lambda) = \tau_0 \times e^{-(\lambda-(\lambda_0+\Delta \lambda_{\rm{abs}}))^2/((\lambda_0+\Delta \lambda_{\rm{abs}})b_{\rm{D}})/c)^2},
\end{equation}
where $\tau_0$, $\lambda_0$, $b_{\rm{D}}$, and $c$ are the central optical depth of the line component, the central wavelength of the line component, the Doppler line width ($\sqrt2\sigma$), and the speed of light. The wavelength offset is related to the velocity offset by $\Delta \lambda_{\rm{abs}} = \Delta v \lambda_0/c$. For the \mgii\ and \feii\ doublets, $\tau_{\rm{0,B}}/\tau_{\rm{0,R}}=$ 2 and 0.2891, respectively \citep{Morton_2003}. We further rewrite $\tau_{\rm{0}}$ using the column density of the ion based on 
\begin{equation}
    N(\rm Mg\ or\ Fe) = \frac{\tau_0 b_D}{1.497\times10^{-15}\lambda_0 f}\  \rm{cm}^{-2},
\end{equation}
where $\lambda_0$ = 2803.5315 (2600.1729) \AA\ and $f$ = 0.3058 (0.239) are the rest-frame vacuum wavelength and oscillator strength of the $\lambda\lambda$2803 (2600) member of the \mgii\ (\feii) doublet \citep{Morton_2003}.

The Gaussian profile for the emission component in the continuum-normalized spectrum is written as
\begin{equation}
    F(\lambda) = 1 + G(A_B,\lambda_B+\Delta \lambda_{\rm{emi}},\sigma) + G(\frac{A_B}{r},\lambda_R+\Delta \lambda_{\rm{emi}},\sigma),
\end{equation}
where $G(A, \lambda,\sigma)$ represent the Gaussian function with amplitude $A$, mean $\lambda$, and standard deviation $\sigma$. The emission amplitude for each line in the doublet is tied by the amplitude ratio $r=A_B/A_R$. For both emission and absorption, we set the width for each line in the doublet to be the same.

\begin{figure*}
\centering
\includegraphics[width=0.495\textwidth]
{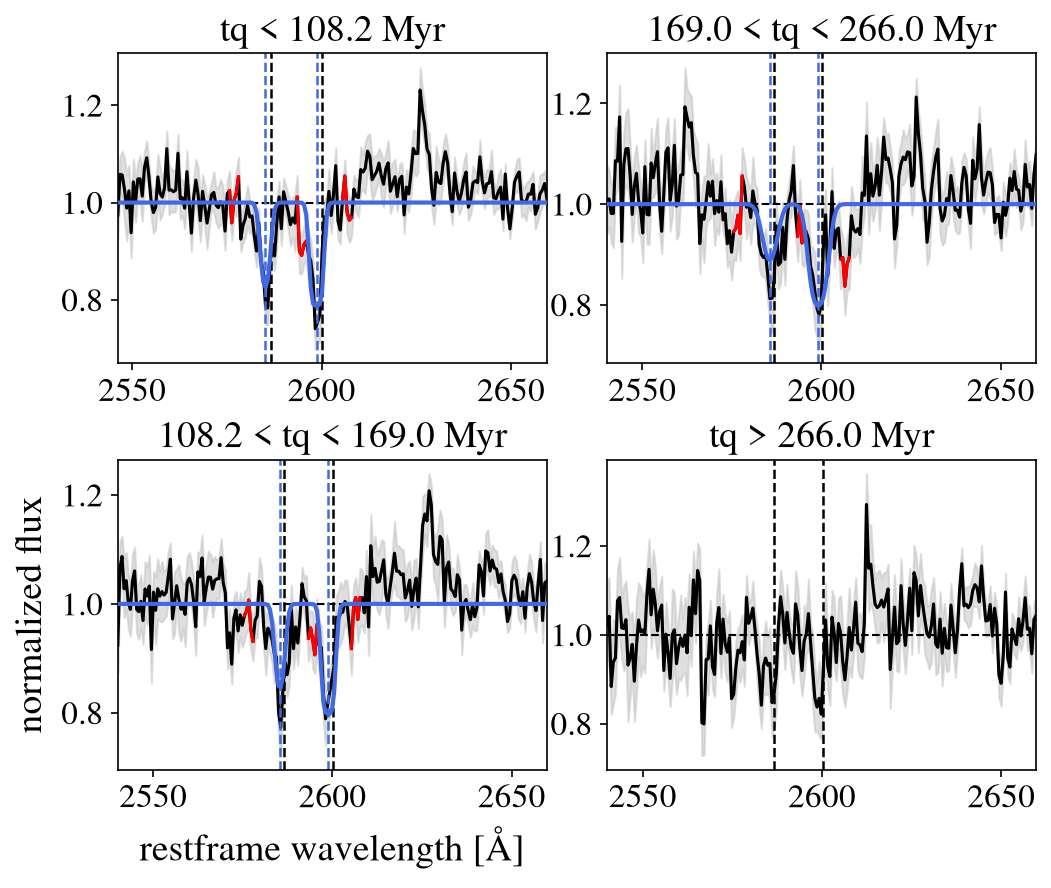}
\includegraphics[width=0.495\textwidth]
{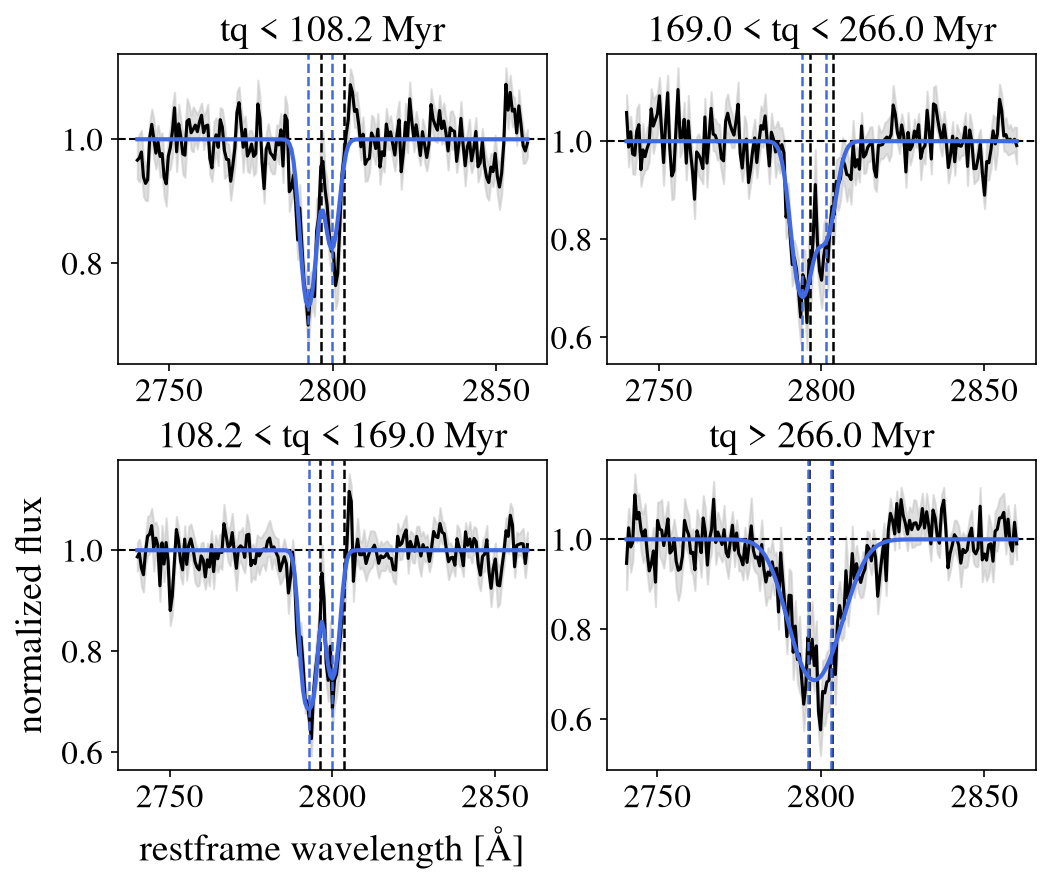}
\caption{\feii\ (left 4 panels) and \mgii\ (right 4 panels) profiles in the $t_q$ stacks. The black curves with gray shaded regions are the continuum-normalized flux and the uncertainty in the stacked spectra. The restframe wavelengths of the lines are indicated by black vertical dashed lines. The best-fit one component absorption models are plotted as blue curves and the line centers of the model indicating outflow velocities are plotted as blue vertical dashed lines. The \feii\ absorption in the largest $t_q$ stack does not pass our quality check to be fitted. \blue{Regions affected by Mn II lines in the \feii\ vicinity which are masked when modeling are marked in red.}
\label{fig:tq_stack}}
\end{figure*}

To account for the variety of spectral profiles, we consider the following four models:
\begin{enumerate}
    \item one-component absorption: $I_{doublet}(\lambda) = I(\lambda)$. The free parameters are $v$, $N$, $C_{f}$, $b_{D}$.
    \item one-component absorption with emission: $I_{doublet}(\lambda) = I(\lambda)F(\lambda)$. The free parameters are $v$, $N$, $C_{f}$, $b_{D}$, $A_B$, $r$, $v_{emi}$, $\sigma$.
    \item two-component absorption: $I_{doublet}(\lambda) = I_{sys}(\lambda)I_{out}(\lambda)$, where $I_{sys}(\lambda)$ is the systemic component with $\Delta v=0$, while $I_{out}(\lambda)$ represents the outflow component. The free parameters are $N_{sys}$, $b_{D,sys}$, $C_{f,sys}$, $N_{out}$, $b_{D,out}$, $C_{f,out}$, and $v_{out}$.
    \item two-component absorption with emission: $I_{doublet}(\lambda) = I_{sys}(\lambda)I_{out}(\lambda)F(\lambda)$. The free parameters are $N_{sys}$, $b_{D,sys}$, $C_{f,sys}$, $N_{out}$, $b_{D,out}$, $C_{f,out}$, $v_{out}$, $A_B$, $r$, $v_{emi}$, and $\sigma$.
\end{enumerate}
We model the \feii\ and \mgii\ doublet profiles separately, fitting \mgii\ lines with all four models, and \feii\ lines with models 1 and 3 as they are less susceptible to emission infilling due to the presence of other Fe non-resonant transitions \citep[e.g.,][]{Zhu_2015}. \blue{We mask the regions around the Mn II $\lambda\lambda$2576.877, 2594.499, and 2606.462 lines near the \feii\ doublet using masks of $\pm2$\AA, $\pm1.5$\AA, and $\pm2$\AA, respectively, to minimize their potential impact on the \feii\ profile modeling.} We then compare the Bayesian Information Criterion (BIC; \citealt{Schwarz_1978}) values for all fitting results and determine the best-fit model as the one with the smallest BIC value. The simplest model (model 1 above) is preferred for the majority \blue{of the fits (66\% for \mgii\ and 100\% for \feii)}. In cases where models other than model 1 are preferred, $\Delta$BIC between the preferred model and model 1 are $<$ 10, indicating weak statistical significance. We therefore adopt results from fitting the one-component absorption model hereafter for consistency. We show the \feii\ and \mgii\ regions with the best-fit one-component absorption models for the four stacks in $t_q$ in Figure \ref{fig:tq_stack}. Spectra and best-fit models for other stacks\blue{, additional fitting details, and discussions on outflow properties derived from other models} are included in Appendix \ref{appendix stack}. 

For each parameter, we take the median of the corresponding posterior distribution from fitting as the final estimated value, and the 16th and 84th percentiles as uncertainties. The velocity of the absorbing gas, in the adopted one-component absorption model, is measured as the velocity offset of the line centers compared to the restframe wavelength. We interpret velocities that remain negative within its uncertainty as indicative of gas outflowing. We adopt the line center offset ($\Delta v$) as the outflow velocity, because it best represents the bulk motion of the absorbing gas along the line of sight. We observe outflows in all stacked spectra except for the one with the largest $t_q$, indicating ubiquitous cool gas outflows in \squiggle\ PSBs. We estimate other outflow properties in the following section. 

\subsection{Derivation of Outflow Properties} \label{sec: outflow derivations}






From the best-fit outflow parameters in Section \ref{sec: line model}, we further estimate the outflow mass $M_{\rm out}$ and mass outflow rate $\dot{M}_{\rm out}$ via 
\begin{align}
M_{\mathrm{out}} &= 1.4\,m_p\,\Omega\,N_{\mathrm{H}}\,R_{\mathrm{out}}^{2}, \\
\dot{M}_{\mathrm{out}} &= 1.4\,m_p\,\Omega\,N_{\mathrm{H}}\,R_{\mathrm{out}}\,v_{\mathrm{out}},
\end{align}
where 1.4$m_p$ is the mean atomic weight corrected for He abundance, $\Omega$, N$_{\rm H}$, $R_{\rm out}$ are the solid angle subtended by the outflow, column density of hydrogen along the line of sight, and the outflow shell radius, respectively. This model assumes a spherically symmetric thin-shell outflow and $\Omega$ is related to the fitted covering fraction via $\Omega = 4\pi C_f$ \cite[e.g.,][]{Rupke_2005b,Davies_2024}\footnote{\blue{In the classic model in the literature, $\Omega = 4\pi C_f C_{\Omega}$, where $C_{\Omega}$ ranging from 0 to 1 is the large-scale covering fraction related to the wind's opening angle. This parameter cannot be measured from the data and is taken to be 1 in our case, as this work analyzes stacked spectra from hundreds of randomly-oriented galaxies.}}.

We convert the ion (Fe and Mg) column density to the hydrogen column density via
\begin{equation}
    \log(N_{\mathrm{H}}) = \log(N_{\mathrm{ion}}) - \chi_{\mathrm{ion}} - \delta_{\mathrm{depl,ion}} 
- \log(\frac{N_{\mathrm{ion}}}{N_{\mathrm{H}}})_{\odot},
\end{equation}
following the literature \citep[e.g.,][]{Rubin_2014,Roberts-Borsani_2019,Valentino_2025}. We assume the ionization fraction $\chi_{\mathrm{ion}}=1$ for Mg and Fe, and adopt the median values in the Galactic disk for the depletion onto dust ($\delta_{\mathrm{depl,ion}}=-1.7$ for Fe, $-0.8$ for Mg; \citealt{Jenkins_2009}). $\log(\frac{N_{\mathrm{ion}}}{N_{\mathrm{H}}})_{\odot}$ represents the solar metallicity and is $-4.49$ for Fe and $-4.42$ for Mg \citep{Savage_1996}.

We do not have spatial information for the outflows in stacked spectra and therefore take the outflow shell radius ($R_{\rm out}$) to be the median of the effective radius (3.46 kpc) for a subsample of \squiggle\ PSBs as measured in \citet{Setton_2022}. This value is within the typical range of 1--5 kpc assumed in similar studies \citep[e.g.,][]{Rupke_2005b,Roberts-Borsani_2019,Davies_2024}. \blue{Uncertainties on $M_{out}$ and $\dot{M}_{out}$ are estimated as the difference between the best-fit value and the 16th percentile of the distribution obtained by applying the same calculations to the full posterior of the outflow parameters. The resulting values are included in Table \ref{tab:outflow_properties}. Although these nominal fitting uncertainties are relatively small, systematic uncertainties arising from assumptions of outflow structures and composition can be much larger, and the saturation of the absorption profiles implies that our mass estimates should be treated as lower limits. We discuss these sources of uncertainty in more detail in the following section.}



\subsection{Uncertainties in Outflow Properties} \label{sec: outflow uncertainties}

While absorption lines provide relatively robust estimates of $v_{out}$, the inferred $M_{out}$ and $\dot{M}_{out}$ are subject to larger uncertainties from multiple underlying assumptions, which we discuss below. 

First, assumptions in converting $N_{ion}$ to $N_{H}$ (equation 7) are largely based on local observations and values. Because the chemical composition, dust content, and depletion of the ion onto dust in the outflowing gas are unknown and difficult to measure, studies including this work typically assume solar metallicity and Milky Way dust depletion measurements for the ions. However, even in the local Galactic ISM, the dust depletion factors span a range of values: $-(0.3-1.5)$ dex for Mg and $-(1.0-2.3)$ dex for Fe \citep{Jenkins_2009}. It is also commonly assumed that all Mg and Fe ions are in the form of \mgii\ and \feii\ since they are likely dominant in photoionized gas of T $\sim10^4$ K \citep{Murray_2007}, but ions in other ionization states might exist as well. Thus adopting single values for these quantities for a population of galaxies is an oversimplification and likely only provides an estimate of the outflow column density to an order of magnitude. \citet{Moretti_2026} recently compared the $N_{H}$ inferred from \feii, \mgii, and \nad\ absorption under typical local assumptions with a direct measurement from Lyman-$\alpha$ absorption in a quiescent galaxy at $z = 2.41$. They found broad agreements between direct measurement and $N_{H}$ from $N_{Na}$ and $N_{Fe}$, and larger discrepancy for $N_{H}$ from $N_{Mg}$, which they attribute to dust depletion differences for Mg.

The geometry of the outflow introduces an additional source of uncertainty. Previous studies have shown that both the detection rate and velocity of absorption-line outflows depend on the inclination of the host galaxy \citep[e.g.,][]{Heckman_2000,Rubin_2014,Sun_2024}. Because absorption lines probe only the line-of-sight component of the gas motion, the measured outflow velocity could underestimate the true velocity. While studies typically make conservative assumptions (1--5kpc) on the outflow radii, outflows can extend from a few to tens of kpc \citep[e.g.,][]{Chen_2010,Rubin_2014,Perrotta_2024} and the outflow radii may change with time as gas moves out. Furthermore, the formalism discussed in Section \ref{sec: outflow derivations} neglects the spatial density profile of the outflowing gas, which would miss possible dense material concentrated in a small velocity range, underestimating the true outflow column density \citep[e.g.,][]{Huberty_2024}.

Finally, the intrinsic line properties combined with mixed physical conditions in outflows can complicate the modeling and interpretation. Although modeling with an emission component does not produce a statistically better fit for our \mgii\ profiles (Section \ref{sec: line model}), emission infilling can still have a non-negligible effect on the line shape and therefore outflow measurements. Moreover, \mgii\ is easily saturated and provides only a lower limit on the outflow column density in such cases. While \feii\ is much less susceptible to emission infilling owing to available \feii* fine structure transitions, it has a smaller oscillator strength and is weaker than \mgii. It has been observed in the literature that different elemental tracers can yield different outflow properties \citep[e.g.,][]{Perrotta_2023,Valentino_2025}. In our sample, $v_{\rm out, Mg}$ is generally $\sim$200 km/s faster than $v_{\rm out, Fe}$, while $N_{\rm H, Mg}$ (therefore derived $M_{out}$ and $\dot{M}_{out}$) is approximately an order of magnitude lower than $N_{\rm H, Fe}$. Although emission infilling can lead to overestimates of $v_{\rm out,Mg}$, \mgii\ is also expected to be more sensitive to more extended, lower-density gas and $|v_{out, Mg}| > |v_{out, Fe}|$ has been observed in multiple studies \citep[e.g.,][]{Erb_2012,Kornei_2012,Prusinski_2021,Perrotta_2023,Kehoe_2025}. \blue{The intrinsic blue-to-red doublet ratio in the optically thin regime is 1:3.46 for \feii\ and 2:1 for \mgii\ based on their oscillator strengths. The doublet ratios observed in all of our stacks indicate saturation, so our $N_{\rm H}$ values represent lower limits only. $N_{\rm H,Fe}$ is more than an order of magnitude larger than $N_{\rm H,Mg}$ in our stacks.} Therefore, consistent with previous findings, \mgii\ appears to more effectively trace higher-velocity gas, whereas \feii\ provides a more stringent constraint on the column density.


In summary, the absolute values of $M_{out}$ and $\dot{M}_{out}$ derived from absorption lines based on standard literature approaches are subject to large systematic uncertainties from various underlying assumptions and provide only order-of-magnitude estimations. $M_{out}$ estimations from lines such as \mgii\ which are particularly susceptible to emission infilling and saturation are even more uncertain. While variations in metallicity, dust depletion and ionization fractions could alter the inferred masses by factors of a few, the combined effect of line saturation, emission infilling, outflow geometry would make the inferred $M_{out}$ and $\dot{M}_{out}$ lower limits. In light of these caveats, we adopt assumptions consistent with common practices in the literature in Section \ref{sec: outflow derivations} to facilitate comparisons with previous studies and assess the relative strength of outflows in intermediate-$z$ PSBs. We use the \feii\ and \mgii\ measurements in tandem in the following sections, focusing on the \feii-based $v_{out}$ and $M_{out}$ and using \mgii\ measurements only to examine $v_{out}$.

\section{Results}

We observe blueshifted \feii\ and \mgii\ absorption (negative measured velocity) in all stacked spectra except the one for the largest $t_q$ group. The addition of a systemic component fixed at $v=0$ does not improve the fitting (Section \ref{sec: line model}; also see Appendix \ref{appendix stack}) and the line profiles are well described by a single blueshifted Voigt-profile doublet. This result suggests that cool gas outflows are ubiquitous in the \squiggle\ sample representing massive intermediate-$z$ PSBs. We further explore the correlation between outflow and galaxies properties, as well as outflow energetics in this section.


\subsection{Outflows and PSBs Properties} \label{sec: outflow properties}

\begin{figure*}
\centering
\includegraphics[width=\textwidth]{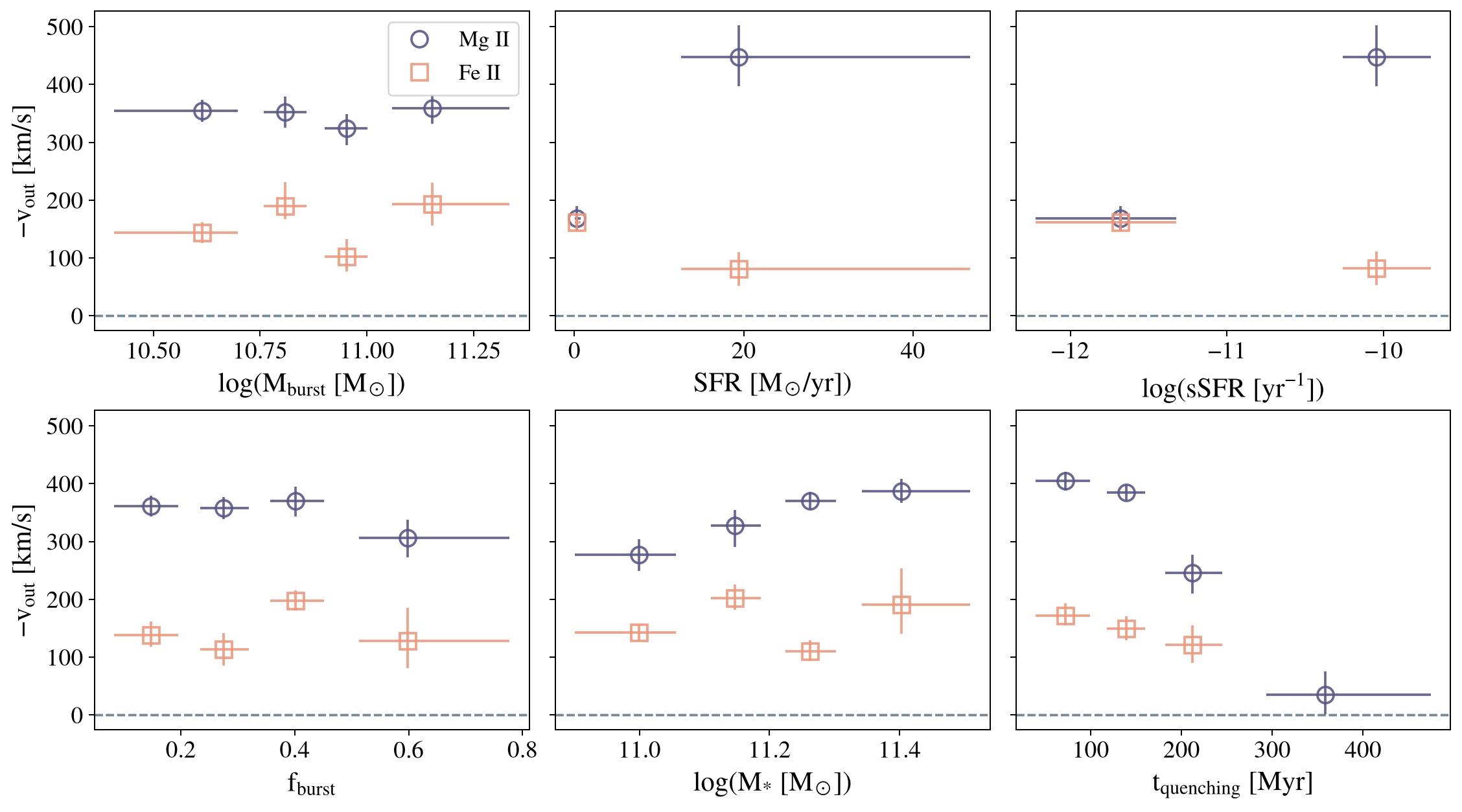}
\caption{Outflow velocity ($v_{out}$) in different stacks binned by galaxy physical and SFH properties, derived from modeling \feii\ and \mgii\ absorption profiles with the one-component absorption model described in Section \ref{sec: line model}. Plotted values and errorbars for the galaxy properties correspond to the median and the 16th–84th percentile range of its distributions within each stack.
\label{fig:stack_vout}}
\end{figure*}

\begin{figure*}
\centering
\includegraphics[width=\textwidth]{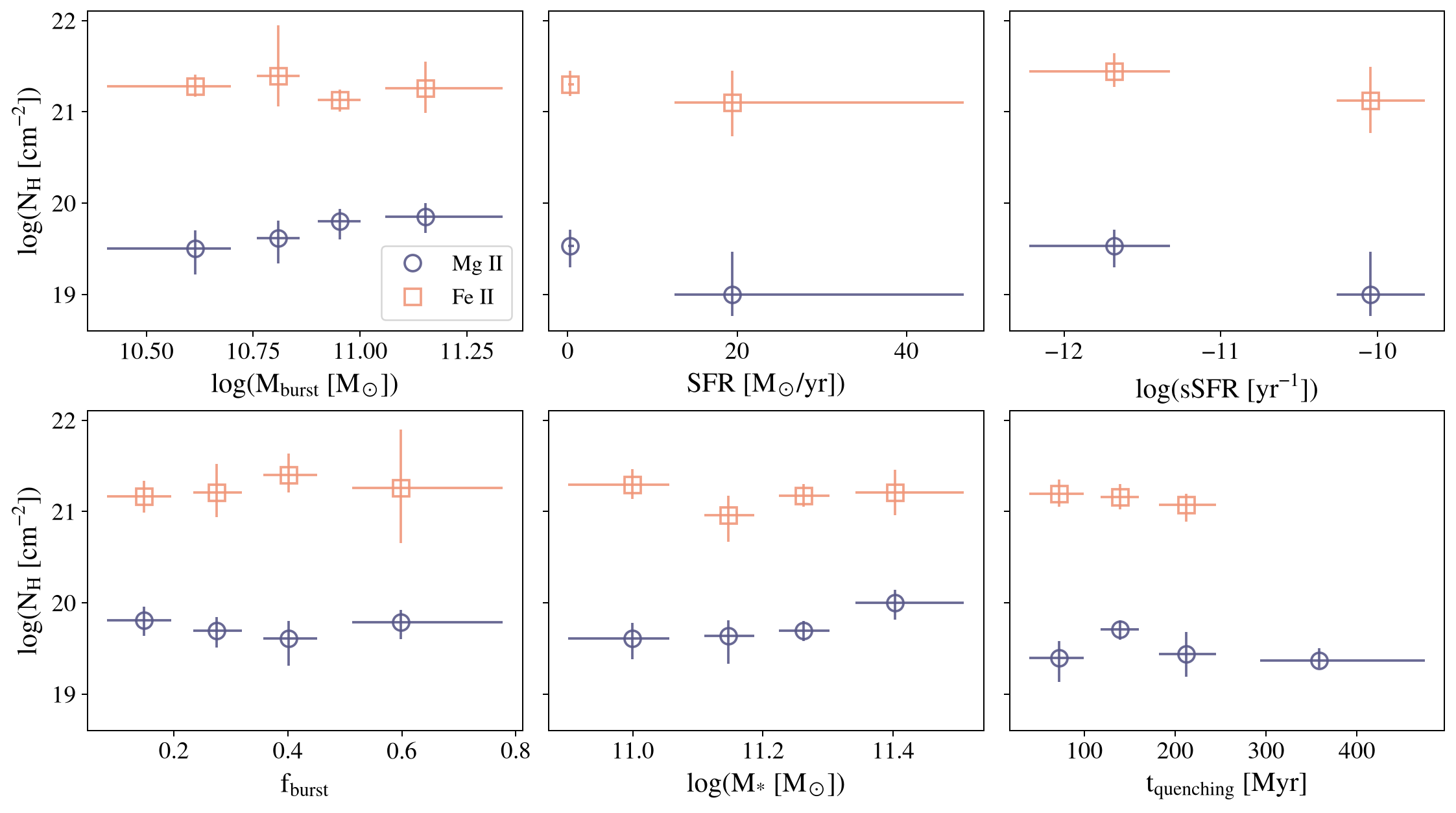}
\caption{Same as Figure \ref{fig:stack_vout} but for the outflow column density N(H). See Section \ref{sec: outflow properties} for more details.
\label{fig:stack_nh}}
\end{figure*}

\begin{figure*}
\centering
\includegraphics[width=\textwidth]{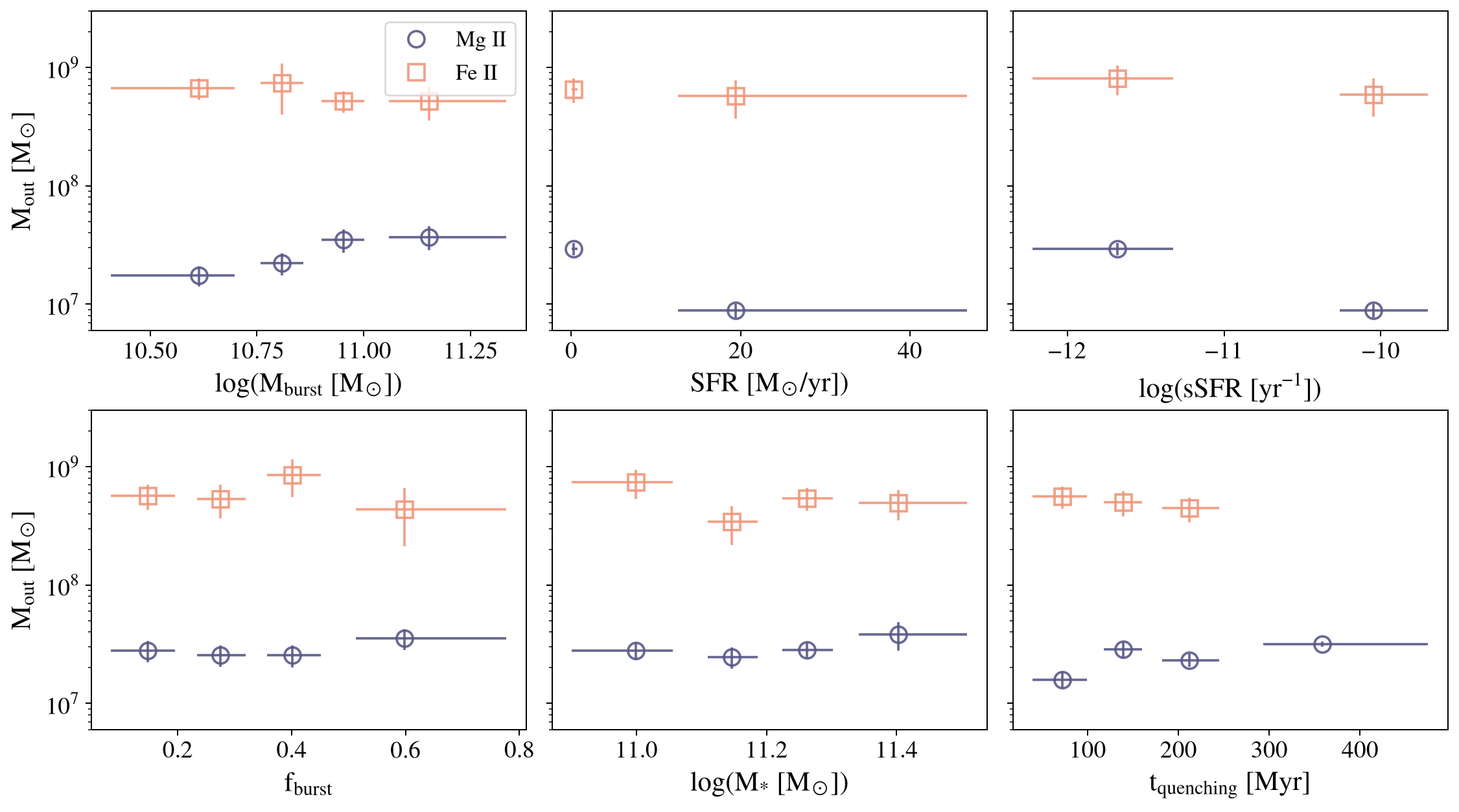}
\caption{Same as Figure \ref{fig:stack_vout} but for the outflow mass ($M_{out}$). See Section \ref{sec: outflow properties} for more details.
\label{fig:stack_mout}}
\end{figure*}

\begin{figure*}
\centering
\includegraphics[width=\textwidth]{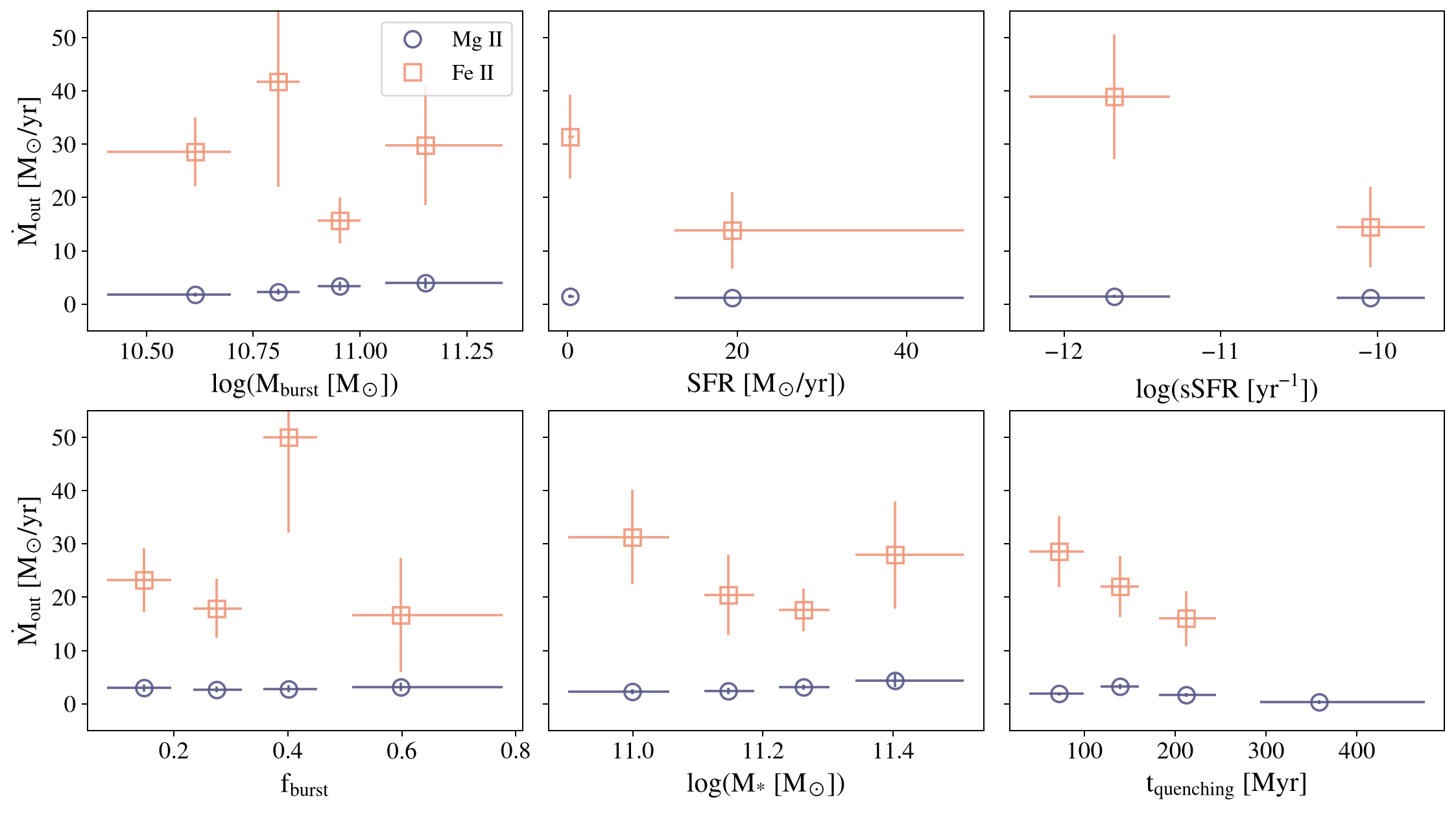}
\caption{Same as Figure \ref{fig:stack_vout} but for the mass outflow rate ($\dot{M}_{out}$). See Section \ref{sec: outflow properties} for more details.
\label{fig:stack_mdot}}
\end{figure*}

In Figures \ref{fig:stack_vout}--\ref{fig:stack_mdot}, we plot derived outflow properties versus the corresponding galaxy properties in which we make different stacks. 

The outflow properties inferred from \mgii\ are largely consistent within each set of stacks and exhibit little correlation with host galaxy properties, with the exception of a relatively more pronounced decreasing trend between $v_{\rm out, Mg}$ and $t_q$. Although $v_{\rm out, Mg}$ is higher in the stack with larger SFR, the limited SFR range spanned by the \squiggle\ sample prevents us from confirming a robust trend. Despite insufficient SNR in the largest $t_q$ stack, $v_{\rm out, Fe}$ appears to show a similar decreasing trend with $t_q$. \blue{In addition, $v_{\rm out, Mg}$ appears to increase with $M_*$, and $\dot{M}_{\rm out, Fe}$ and $v_{\rm out, Fe}$ decreases with SFR and sSFR. Because $M_{\rm out}$ remains largely consistent within each set of stacks, the observed declines in $\dot{M}_{\rm out, Fe}$ with increasing (s)SFR are likely driven by the corresponding decrease in $v_{\rm out, Fe}$. These trends are likely tentative as they are observed only in one of the tracers and we further compare them to outflow scaling relations below.} In summary, the most prominent trend is the decline of $v_{out}$ with increasing $t_q$, which, if interpreted as an evolutionary sequence, suggests that the observed outflows may be relic outflows launched at earlier times and have been slowing down since then. Other studies have similarly found that the highest-velocity outflows tend to be associated with the youngest stellar populations \citep[e.g.,][]{Davis_2023,Sun_2024,Perrotta_2024}. 





\begin{figure*}
\centering
\includegraphics[width=\textwidth]{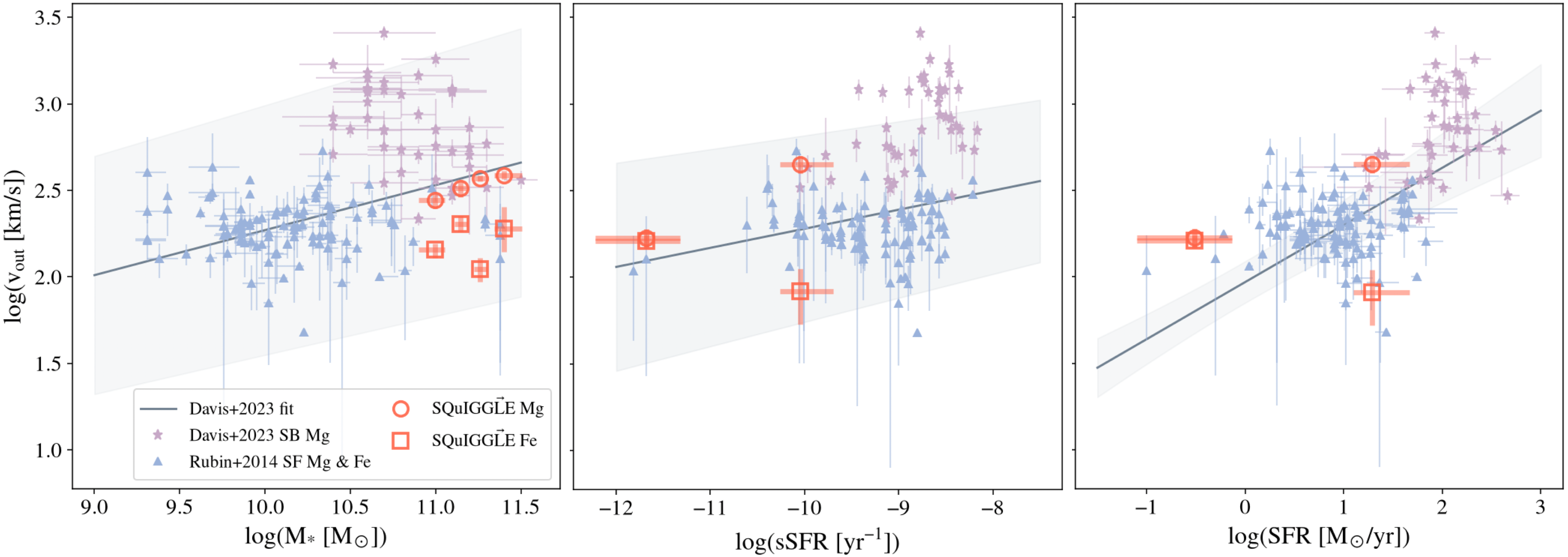}
\caption{Outflow scaling relations based on a compilation of large literature samples in \citet{Davis_2023}. The relations shown correspond to their fits using $V_{\rm avg}$ in their Table 2. The star-forming galaxies ($0.3<z<1.4$) from \citet{Rubin_2014} and starburst galaxies ($0.4<z<0.8$, the HizEA sample) from \citet{Davis_2023} are included for reference. Outflows in \squiggle\ PSBs are in general consistent with the scaling relations commonly observed in star-forming galaxies.
\label{fig:scaling}}
\end{figure*}

Velocity is the best constrained outflow property in our analysis and we further place our results in the context of the literature by comparing the outflow velocities in \squiggle\ PSBs to known outflow scaling relations. \citet{Davis_2023} constructed scaling relations between $v_{out}$ and SFR, M$_*$, and sSFR from literature samples of star-formation driven outflows traced by near-UV absorption lines (Mg, Fe, or Si). To construct a reasonably homogeneous comparison sample and minimize the effects of redshift evolution, \citet{Davis_2023} limited their compilation to studies of ions with similar ionization potentials over the range $0 < z < 1.5$. In Figure \ref{fig:scaling}, we plot the \squiggle\ measurements and the best-fit scaling relations from \citet{Davis_2023}. For comparison, we also include the HizEA starburst galaxies from \citet{Davis_2023} and the star-forming galaxies from \citet{Rubin_2014}, which lie at similar redshifts to the \squiggle\ sample. All velocity measurements and scaling relations shown in Figure \ref{fig:scaling} are based on line-centroid velocities, consistent with our methodology. The outflows in \squiggle\ PSBs are largely consistent with the scaling relations. However, $v_{out}$ in the lower SFR stack appears somewhat elevated, raising the question of whether the current SF is sufficient to drive the outflows. We look further into the outflow energetics and discuss possible outflow driving mechanisms in the following sections.

\subsection{Outflow Energetics and Comparison to Current SF and AGN} \label{sec: outflow energetics}

\begin{figure}
\centering
\includegraphics[width=\columnwidth]{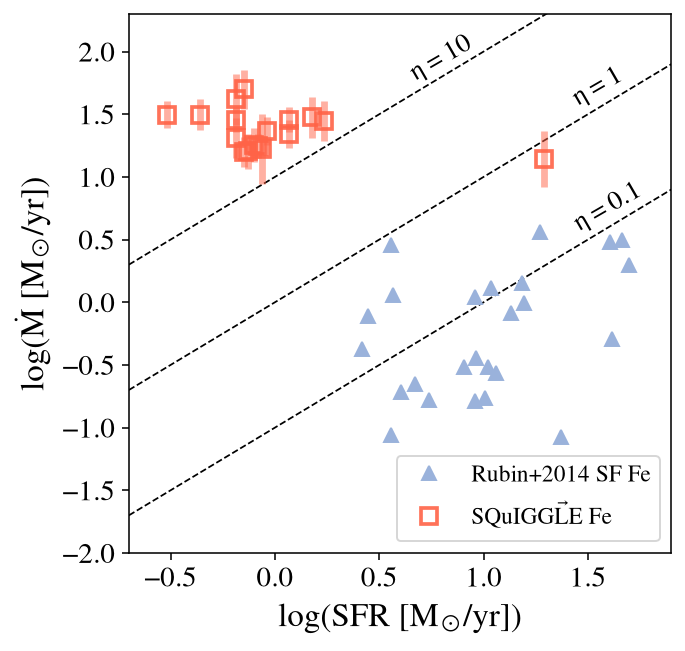}
\caption{The mass outflow rates estimated using \feii\ plotted against the median SFR in each stack. Similar measurements of star-forming galaxies ($0.3<z<1.4$) in \citet{Rubin_2014} are plotted for reference (only those with log$\dot{M}>-2$ are shown, and these estimations are lower limits based on conservative assumptions according to the paper). \squiggle\ PSBs show large mass loading factors of $\gtrsim10$ similar to recent findings of neutral outflows in $z\gtrsim2$ rapidly quenched galaxies.
\label{fig:mdot_sfr}}
\end{figure}




We plot in Figure \ref{fig:mdot_sfr} $\dot{M}_{out}$ from \feii\ measurements versus the median SFR in each stack. Outflows in our stacks have large mass loading factors ($\eta=\dot{M}_{out}/\rm{SFR}$) of \blue{$\sim1-120$}, exceeding those found in star-forming galaxies and local luminous infrared galaxies ($\eta\sim0.1-1$; e.g., \citealt{Rupke_2005b,Rubin_2014,Roberts-Borsani_2019}). We note that our estimated $\eta$ are lower limits because most of the SFRs in our sample are upper limits (Section \ref{sec: sample}; see details in \citealt{Suess_2022}) \blue{while $\dot{M}_{out}$ are lower limits (Section \ref{sec: outflow derivations})}.

AGN are often invoked to drive outflows with large mass loading factors and here we examine whether the presence of a current AGN is related to our observed outflows. There are $\sim5$\% AGN identified via optical emission lines (the mass-excitation diagram) in the \squiggle\ sample \citep{Greene_2020}. We make stacked spectra for galaxies with (isAGN = 1) and without (isAGN = 0) identified AGN to explore whether outflows are correlated with AGN. The continuum-normalized spectra in Figure \ref{fig:agn_stack}, Appendix \ref{appendix agn stack} indicates that the ISM \feii\ and \mgii\ absorption in the AGN stack does not satisfy the quality check in Section \ref{sec: line model} (either EW$<1$~\AA\ or SNR$_{\rm abs}<4$), whereas the no-AGN stack shows blueshifted absorption similar to those seen in other stacks. The no-AGN stack has a higher SNR because it includes substantially more galaxies than the AGN stack, which could potentially affect absorption-line detectability. To assess this effect, we randomly draw the same number of galaxies as in the AGN stack from the no-AGN sample to generate 100 reduced-size no-AGN stacks and measure the \feii\ and \mgii\ absorption after subtracting the stellar continuum. \blue{97}\% and \blue{40}\% of these reduced non-AGN stacks satisfy EW$> 1$~\AA\ and SNR$_{\rm abs}>4$ for the \mgii\ and \feii\ lines, respectively. Therefore, despite its smaller sample size, the AGN stack should be capable of recovering ISM absorption if present, at least for the \mgii\ line. While this result suggests that cool gas outflows do not preferentially occur in \squiggle\ PSBs currently hosting AGN, potential \mgii\ emission from AGN host galaxies \citep[e.g.,][]{Weiner_2009} may partially infill the absorption features, making it difficult to draw a firm conclusion.


We further estimate the energy and momentum that could be provided by current SF via supernovae and compare them to those of the outflows. We estimate the momentum and energy injection rates from supernovae following \citet{Veilleux_2005}:
\begin{align}
    \dot{p}_{\rm SFR}\, [\rm dyne] = 5 \times 10^{33}\, \frac{\rm SFR}{M_{\odot}\ yr^{-1}},\\
    \dot{E}_{\rm SFR}\, [\rm erg\,s^{-1}] = 7 \times 10^{41}\, \frac{\rm SFR}{M_{\odot}\ yr^{-1}},
\end{align}
assuming 10$^{51}$ ergs per supernova per 100$M_{\odot}$ of new stars formed. The outflow momentum and energy rates are calculated as $\dot{p}_{\rm out} = \dot{M}_{\rm out}v_{\rm out}$ and $\dot{E}_{\rm out} = \frac{1}{2}\dot{M}_{\rm out}v_{\rm out}^2$ using Fe measurements.

The ranges for $\dot{p}_{\rm out}/\dot{p}_{\rm SFR}$ and $\dot{E}_{\rm out}/\dot{E}_{\rm SFR}$ are \blue{0.1$-$26} and \blue{0.002$-$1.5}, respectively. These ratios are lower limits because our SFRs are upper limits and not all $\dot{p}_{\rm SN}$ and $\dot{E}_{\rm SN}$ would be coupled to outflows. Accounting for the efficiency of energy transfer to the ISM, typically 0.1 in the literature \citep[e.g.,][]{Murray_2005,Thompson_2024}, $\dot{E}_{\rm out}/\dot{E}_{\rm SFR}$ in our sample could be an order of magnitude larger. Thus energy from current SF is insufficient to support the observed outflows. We discuss the possibility of outflows driven by past SF and/or AGN events in the following section.

\section{Discussion} \label{sec: discussion}

\subsection{Outflows Driven by Past SF or AGN} \label{sec: past sf/agn}

Recent studies reporting neutral outflows with large $v_{out}$ and $\eta$ have proposed that these outflows may be relics launched by past SF and/or AGN activity \citep[e.g.,][]{Taylor_2024,Davies_2024,Taylor_2026}. \squiggle\ PSBs have experienced recent starbursts with past peak SFRs $\sim$ a few hundred $M_{\odot}$ yr$^{-1}$ (\citealt{Suess_2022}, Setton et al. 2026, in prep). Assuming the past peak SFRs ($\sim100\times$ current SFR) would decrease the estimated $\dot{p}_{\rm out}/\dot{p}_{\rm SN}$, $\dot{E}_{\rm out}/\dot{E}_{\rm SN}$, and $\eta$ in Section \ref{sec: outflow energetics} by two orders of magnitude to $<1$, in agreement with the SF-driven outflow scenario.

Even though we do not detect outflows in the AGN stack and only a small fraction of \squiggle\ PSBs show signs of current AGN activity, AGN could have been on at an earlier time and contributed to driving the outflows. Here we test whether AGN could provide enough energy for the observed outflows, approximating the bolometric luminosity of past AGN using that of current ones observed in the \squiggle\ sample. This approximation is conservative since the color criteria for \squiggle\ PSBs exclude optically luminous AGN (e.g., quasar-like objects) and more luminous AGN often accompany starbursts \citep[e.g.,][]{Wild_2010}. We take the bolometric correction to be 600 \citep[e.g.,][]{Kauffmann_2009,Netzer_2009}, $L_{bol} = 600\times L_{\rm [O\ III]}$, and $L_{\rm [O\ III]}$ to be $5\times 10^{41}$ erg/s, roughly the median value in \citet{Greene_2020}. Following \citet{Davies_2024}, we assume energy conserving AGN-driven outflows and estimate the AGN energy injection rate as $\dot{E}_{\rm AGN} = 0.05L_{bol}$ (5\% of the AGN bolometric luminosity; \citealt{King_2015} and references therein), and the momentum injection rate as $\dot{p}_{\rm AGN}$ = $L_{bol}/c$. We find $\dot{p}_{\rm out}/\dot{p}_{\rm AGN} \lesssim \blue{6.2}$ and $\dot{E}_{\rm out}/\dot{E}_{\rm AGN} \lesssim0.04$. In energy-conserving outflows, $\dot{p}_{\rm AGN}$ could be boosted by a factor of $\sim5-$20 due to entrainment of ISM gas in the wind \citep[e.g.,][]{Faucher_2012}. 

While our results suggest that the observed outflows could be driven by past starbursts and/or past AGN with luminosities comparable to those currently observed in \squiggle\ PSBs, connecting present outflows to past events is complicated by the different timescales involved. Models predict that the mechanical energy released by stellar winds and supernovae can persist for a few tens of Myr after a starburst \citep[e.g.,][]{Leitherer_1999}, and observational analysis on dwarf galaxies indicates that starburst-driven outflows have a short lifetime of $\lesssim25$ Myr after the decline of the starburst \citep{McQuinn_2018}. The outflow in the stack of $t_q>100$ Myr, if launched by the previous starburst, may require energy input from other sources after the galaxies are quenched. On the other hand, AGN can vary on timescales shorter than changes in SF in galaxies \citep[e.g.,][]{Novak_2011}. \citet{Krishna_2025} recently found a significant excess of PSBs among quasars compare to star-forming galaxies and that these quasars are preferentially hosted by massive galaxies ($\sim10^{11.3}M_{\odot}$). It is possible that many \squiggle\ PSBs, which are similarly massive, experienced a recent quasar phase that has since faded \citep[e.g.,][]{French_2023}. Under this scenario, the outflows could have been initially launched by the previous starburst and/or quasar events and subsequently maintained by lower-luminosity, episodic AGN activity, consistent with scenarios proposed in similar studies \citep{Almaini_2025,Taylor_2026}. The lack of clear positive correlations between outflow velocities and starburst properties ($f_{burst}$, $M_{burst}$), together with the short lifespan of starburst-driven outflows, might suggest that AGN play a more important role in driving outflows in PSBs. However, additional studies focusing on AGN-host PSBs utilizing outflow tracers less affected by AGN are needed to test this hypothesis.


\subsection{Fate of the Outflowing Gas} \label{sec: outflow fate}


The escape velocity required for gas to leave the ISM can be estimated using the stellar mass and half light radius of the galaxy. Given the narrow ranges of these parameters for \squiggle\ galaxies, we calculate the typical ISM escape velocity using the median $M_*$ (10$^{11.20}$ $M_{\odot}$) and $r_{\rm eff}$ (3.46 kpc) of the sample, $v_{\rm esc, ISM} = \sqrt{\frac{2GM*}{2r_{\rm eff}}} \sim$ 440 km/s, where we assume the radius for the extent of the ISM to be 2$r_{\rm eff}$. We also estimate the CGM escape velocity, above which gas could leave the halo of the galaxy and become gravitationally unbound to the galaxy, as $v_{\rm esc, CGM} = 3v_{\rm circ}$ \citep[e.g.,][]{Veilleux_2020}. We estimate the circular velocity as $v_{\rm circ} = \sqrt{2}S$, where the kinematic parameter $S$ is $log(S) = 0.29\, log(M_*)-0.93$ following the relation of low redshift star-forming galaxies \citep[e.g.,][]{Heckman_2015,Simons_2015}. Using the median $M_*$ of the \squiggle\ sample, we estimate $v_{\rm esc, CGM}\sim$880 km/s, similar to that inferred for galaxies around cosmic noon \citep[e.g.,][]{Taylor_2024,Davies_2024} and is much larger than the $v_{\rm esc, ISM}$.

While $v_{out}$ defined by the line center offset is smaller than $v_{\rm esc, ISM}$ for all stacks, the maximum velocity $v_{max} = v_{out} - 2\sigma$ characterizing the highest-velocity component exceeds $v_{\rm esc, ISM}$ for all $v_{max,Mg}$ and \blue{20}\% of $v_{max,Fe}$, with a few cases even slightly exceeding $v_{\rm esc, CGM}$. This result indicates that the observed outflows in \squiggle\ PSBs have the potential of displacing some fraction of the outflowing gas into, and occasionally out of, the CGM of the galaxies. Our finding is consistent with the stronger \mgii\ absorptions observed out to $\sim$500 kpc into the CGM of $M_*>10^{11}M_{\odot}$, $0.4 \lesssim z \lesssim 0.8$ PSBs relative to comparison samples of star-forming and quiescent galaxies in \citet{Harvey_2025}. Their observations indicate an excess of extended cool gas in the CGM of PSBs, which they speculate to be related with cool gas outflows. Similarly, extended low-density CGM gas has been observed via \mgii\ and [\ion{O}{2}] emission in the the HizEA sample \citep[e.g.,][]{Rupke_2019,Perrotta_2024}, some of which are found to host $\gtrsim$1000 km/s outflows in \mgii\ and \feii\ absorption \citep{Davis_2023,Perrotta_2023}. The HizEA sample consists of galaxies originally selected as E+A post-starburst galaxies, but later revealed to host compact buried starbursts \citep[e.g.,][]{Tremonti_2007,Diamond-Stanic_2021,Davis_2023}. They have similar masses and redshifts to \squiggle\ PSBs and some show signs of quenching in their SFH \citep{Perrotta_2024}, suggesting that they may represent an earlier evolutionary stage of \squiggle\ PSBs. We further explore outflows in PSBs studied in the literature across different redshifts and relate our results to the context in the following sections.

\subsection{Outflows in PSBs across Cosmic Time} \label{sec: outflows in PSBs}

\begin{figure*}
\centering
\includegraphics[width=\textwidth]{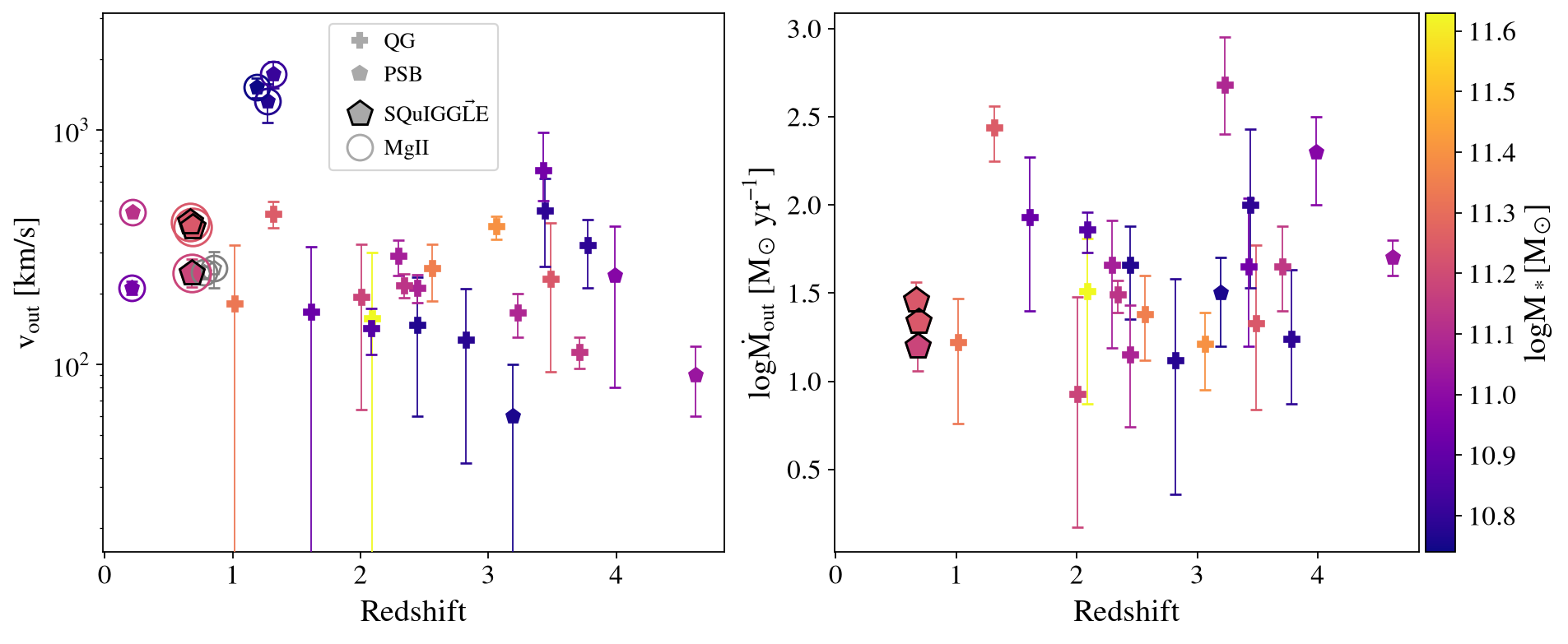}
\caption{Outflow velocities (left) and mass outflow rates (right) plotted against redshift for the $t_q$ stacks in this work (\squiggle\ PSBs; larger symbols with black outlines) and literature studies on PSBs and QGs discussed in Section \ref{sec: outflows in PSBs}. Data points are color coded by host galaxy stellar masses where available (median stellar masses for stacks). We only plot galaxies with $M_*>10^{10.6} M_{\odot}$ for a closer comparison with the \squiggle\ sample (two galaxies from \citealt{Coil_2011} with no available $M_*$ are still included as open markers since they lie at similar redshifts to \squiggle\ PSBs and were analyzed using the same outflow tracer). \blue{For \squiggle\ PSBs, the plotted $v_{out}$ ($\dot{M}_{out}$) is from \mgii\ (\feii) measurements. Literature measurements plotted are either based on \mgii\ or \nad\ lines, with those based on \mgii\ marked by an additional circle in the left panel, as for the \squiggle\ PSBs.} Cool gas outflows in PSBs/QGs have moderate velocities below the galaxy escape velocities in the majority of the cases, and have large $\eta>1$ due to their low current SF. \blue{Literature measurements included in this plot are from \citet{Coil_2011,Maltby_2019,Taylor_2024,Davies_2024,Taylor_2026,Zhu_2026,Sun_2026} and} the data behind this figure is available \blue{as a table} online.
\label{fig:vout_z}}
\end{figure*}

Absorption-line-traced outflows have been widely observed in PSBs across cosmic time. \citet{Coil_2011} found outflows traced by blueshifted \feii\ and \mgii\ absorption in 13 PSBs at $0.2<z<0.8$, with typical velocities\footnote{Outflow velocity hereafter refers to that derived from offset in the line centroid, unless otherwise specified.} of 200$-$300 km/s. They reached a similar conclusion that while some gas may escape to the halo of the galaxy, the bulk of the gas remains bound. At $1<z<1.5$, stacks of PSBs show blueshifted \mgii\ absorption signaling outflows $\gtrsim$1000 km/s (\citealt{Maltby_2019,Taylor_2024}). These outflows have a velocity greater than the escape velocity of the galaxy and neither work found AGN evidence in the optical spectra of their PSB samples. They concluded that such high-velocity outflows are typical in massive ($M_* \gtrsim 10^{10} M_{\odot}$) PSBs at this epoch and might be the residual signature of the feedback process that quenched these galaxies. \mgii\ absorption traced outflows have also been found in two recently quenched galaxies at much higher redshifts of $z\sim4$ and 7, although the velocity is much smaller at $\sim180$ km/s \citep{Wu_2025,Valentino_2025}. The two systems show vastly different mass loading factors, with the $z\sim7$ system having $\eta\sim40$ suggesting an additional ejective mechanism such as an undetected AGN.

\nad\ absorption is another common outflow tracer that typically trace cooler and denser neutral gas than \feii\ and \mgii\ due to its lower ionization potential. In the local universe ($z\lesssim 0.3$), \citet{Sun_2024} studied the \nad\ traced outflows in samples of star-forming galaxies (SFGs), PSBs, and quiescent galaxies (QGs) without AGN and found a trend of decreasing outflow velocity with time since the starburst ended. The outflow velocity in their PSBs is mostly $\lesssim250$ km/s and only in a few cases the gas could escape the galaxy halo. They concluded that it is unclear whether or how these outflows regulate the SF in their host galaxies. On the other hand, studies of low-$z$ PSBs hosting AGN revealed neutral outflows with $\gtrsim1000$ km/s from the \nad\ profile, expelling gas at $\gtrsim10$ M$_{\odot}$/yr \citep{Baron_2020,Baron_2022,Baron_2024}. At $z\gtrsim2$, studies have shown that neutral outflows of $v\gtrsim100$ km/s and $\eta>1$ are prevalent in recently quenched and quiescent galaxies \citep{Davies_2024,Belli_2024,DEugenio_2024,Taylor_2026,Zhu_2026,Sun_2026}. Although the mass loading factors can sometimes be as high as $\gtrsim100$ in these galaxies, the bulk of the outflowing material will remain in the galaxy halo which might later be involved in fountain-like recycling.

We plot $v_{out}$ and $\dot{M}_{out}$ versus redshift from aforementioned studies and from our analysis in Figure \ref{fig:vout_z}, color coded by $M_*$.  We include only galaxies with $M_*>10^{10.6}M_{\odot}$ to facilitate a closer comparison with the massive \squiggle\ PSBs. In addition to following the literature scaling relations for SFGs (Figure \ref{fig:scaling}), outflow properties in \squiggle\ PSBs are consistent with those of similar outflows in PSBs/QGs at comparable and higher redshifts. Apart from several extreme cases of ultra-fast outflows, cool gas outflows in PSBs from the local universe to beyond cosmic noon generally appear to have moderate velocities below the escape velocities of their host galaxies. $\dot{M}_{out}$ as shown in the right panel of Figure \ref{fig:vout_z} does not exhibit a significant redshift evolution either, although we note the caveat that derived $\dot{M}_{out}$ generally have large uncertainties (Section \ref{sec: outflow uncertainties}). 




\subsection{Implications for Quenching}

\begin{figure}
\centering
\includegraphics[width=\columnwidth]{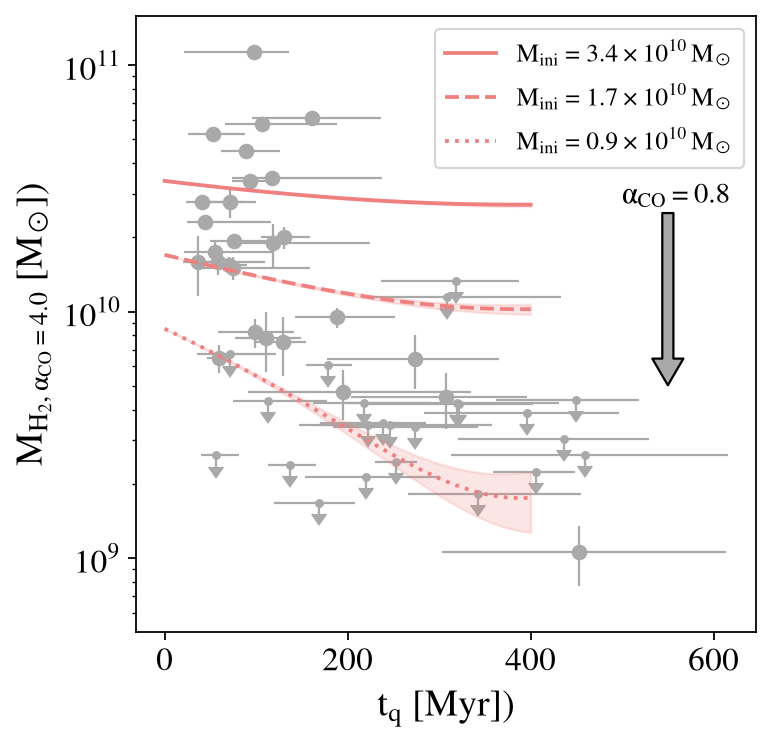}
\caption{$M_{\rm H_2}$ versus $t_q$ for the 50 \squiggle\ PSBs studied in \citet{Setton_2025}. Non-detections are 3$\sigma$ upper limits plotted as small downward arrows. The red curves show the expected evolution of $M_{\rm H_2}(t)$ for gas reservoirs with different initial masses undergoing depletion at a rate of $\dot{M}_{\rm out,Fe}(t)$ based on this work. The big arrow on the right of length 0.7 dex indicates the amount of shift in the background gray points if $\alpha_{CO}=0.8$ instead. Outflows may fully explain the observed trend of decreasing $M_{\rm H_2}$ over time for cases with relatively low molecular gas masses.
\label{fig:mh2_tq}}
\end{figure}

One important question in studying outflows in PSBs is how these outflows connect to the rapid quenching. In the local universe ($z\lesssim 0.3$), the role of outflows in quenching is unclear or often considered minor \citep[e.g.,][]{Sun_2024,Fodor_2025} given the small velocity and mass outflow rate. At $z\gtrsim2$, while outflows are still often unable to remove material out of the halo of the galaxies (Section \ref{sec: outflows in PSBs}), their large mass loading factors (1 to a few hundreds) lead to the conclusion that they could have significant impact on the gas reservoir and may represent a major pathway for rapid quenching \citep[e.g.,][]{Davies_2024,Park_2024,Zhu_2026}. The mass loading factors in \squiggle\ PSBs from \feii\ measurements are \blue{$\sim1-120$}, similarly large to those found for higher redshift PSBs mentioned in the previous section, indicating that outflows may also be an important avenue for fast quenching at $0.5<z<0.9$. Observations at similar redshifts of cool gas in the CGM of PSBs \citep{Harvey_2025} and starburst galaxies (the HizEA sample; e.g., \citealt{Perrotta_2024}) further agree with the quenching picture that outflows are relocating the gas to the galaxy halo. Interestingly, \squiggle\ PSBs lie close to the HizEA galaxies with slower $v_{out}$ and larger $M_*$ in Figure \ref{fig:scaling}, consistent with our speculation that the outflows in \squiggle\ PSBs are relics and with the possibility that \squiggle\ PSBs are coeval descendants of HizEA galaxies. 

\citet{Setton_2025} presented CO(2-1) observations of 50 \squiggle\ PSBs and found a decreasing trend of the inferred molecular gas mass (M$_{\rm H_2}$) with $t_q$ (see also \citealt{French_2018, Bezanson_2022} for similar age trends). This trend, if taken as an evolutionary sequence, implies that galaxies would fully deplete their molecular gas within $\sim250$ Myr of quenching. \citet{Setton_2025} showed that even a combination of lower CO-to-H$_2$ conversion factor ($\alpha_{CO}$) and higher SFRs\footnote{Maximal SFRs based on UV-to-FIR spectral energy distribution fitting with the assumption that birth clouds can have significantly higher dust columns than the ambient ISM. See \citet{Setton_2025} for more details in modeling.} still produces depletion time an order of magnitude larger than expected ($\sim1$ Gyr for some of the galaxies). They speculated outflows as a possible means of gas removal. $\dot{M}_{out}$ from \feii\ measurements in this work roughly decreases linearly with $t_q$ (Figure \ref{fig:stack_mdot}). We model $\dot{M}_{\rm out,Fe}(t)$ as a linear function and compute the corresponding evolution of the molecular gas mass, $M_{\rm H_2}(t)$, for several assumed initial gas masses: $M_{\rm ini}=1.7\times10^{10}M_\odot$ (the median of the detections in \citealt{Setton_2025}) and two additional values offset by $\pm0.3$ dex. The resulting tracks are overplotted with the \citet{Setton_2025} observations in Figure \ref{fig:mh2_tq}. The exact gas depletion time depends on $M_{\rm ini}$ and Figure \ref{fig:mh2_tq} shows that for relatively low $M_{\rm ini}$ values outflows alone may account for the observed decline of $M_{\rm H_2}$ with increasing $t_q$. Such a scenario could arise if $M_{\rm H_2}$ were derived using a ULIRG-like $\alpha_{CO}=0.8$ instead of Milky Way-like $\alpha_{CO}=4.0$ (as illustrated by the arrow in Figure \ref{fig:mh2_tq}), although \citet{Setton_2025} found no data-driven evidence strongly supporting this modification.


We note that outflows studied in this work are in the neutral/weakly-ionized rather than the molecular phase, and it is also possible that the trend in \citet{Setton_2025} does not represent an evolutionary sequence (see more discussions therein). Nevertheless, it remains possible that molecular gas outflows exist but are very challenging to detect (see \citealt{Veilleux_2020} for a review, and references therein), and studies have suggested that $M_{\rm out}$ in the molecular phase is in general larger than, or at least comparable to that in the neutral phase \citep[e.g.,][]{Veilleux_2005,Fluetsch_2019}. Molecular gas could also be converted to the neutral phase (e.g., by shocks/turbulence; \citealt{Hollenbach_1989}) and then carried away by cool gas outflows.

In summary, the combination of outflow velocities exceeding the ISM escape velocity, large mass loading factors, and estimated gas depletion timescale consistent with observations implies that outflows traced by \feii\ and \mgii\ absorption in \squiggle\ PSBs have the potential for quenching SF in these massive ($\gtrsim10^{11}M_{\odot}$) galaxies, possibly by displacing the gas into the CGM and affecting the gaseous environment around the galaxies. Future deep observations of the dense molecular gas phase, the direct fuel for SF, will provide crucial information to build a complete picture of the multiphase gas properties in PSBs. Spatially-resolved observations of the CGM of PSBs will help connect outflow properties and SFH (e.g., time since quenching) to the gas content and distribution in the CGM, offering an important test of the outflow quenching scenario and key insights into the ultimate fate of the gas.

\section{Summary} \label{sec: summary}


While the role of outflows in SF quenching remains ambiguous in the local universe due to small velocity and/or low mass outflow rate, recent studies of neutral outflows in rapidly quenched and quiescent galaxies beyond cosmic noon suggest that outflows may represent a major pathway for rapid quenching. In this work, we search for cool gas outflows and characterize their properties in massive ($M_*>10^{11}M_{\odot}$) PSBs at intermediate redshift of $0.5<z<0.9$, exploring their correlation with galaxy properties, possible driving mechanisms, and connection to quenching. Our primary findings are summarized below:

\begin{enumerate}
    \item Utilizing the \squiggle\ survey containing $>$1300 PSBs, we stack spectra over bins of galaxy physical and SFH properties and study the cool gas outflows traced by \feii\ $\lambda\lambda$2586,2600 and \mgii\ $\lambda\lambda$2796,2803 absorption. We observe blueshifted line centers for both ions in all stacked spectra binned by $M_*$, $M_{burst}$, SFR (sSFR), $f_{burst}$, and $t_q$, except in the bin with the largest $t_q$ ($> 266$ Myr). The line profiles are well described by a single blueshifted Voigt-profile doublet without the need for a systemic component fixed at $v=0$. Our results indicate that cool gas outflows are ubiquitous in massive PSBs at intermediate redshifts.
    
    \item We observe decreasing $v_{out}$ with increasing $t_q$ from both \feii\ and \mgii\ measurements, suggesting that observed outflows might be relics launched at earlier epochs and have since been slowing down. There are no other significant correlations between outflow properties ($v_{out}$, N$_{\rm H}$, $M_{out}$, $\dot{M}_{out}$) and galaxy properties in which we stack. The measured $v_{out}$ values are also broadly consistent with literature scaling relations with SFR, $M_*$, and sSFR.
    
    
    \item The inferred outflow energy and momentum injection rates exceed those expected from current SFR but could be explained by the peak SFR during the past starburst phase. Although we do not detect outflows in the stacked spectrum of galaxies currently hosting AGN, the observed AGN have sufficient energy to support the outflows. Therefore, the outflows could have been launched by the previous starburst and/or AGN events, and may be further sustained over 200 Myr by potential lower-luminosity, episodic AGN activity after quenching.
    
    
    \item The maximum outflow velocity \blue{exceeds the ISM escape velocity in all \mgii\ stacks and, in some cases, even exceeds the CGM escape velocity}, indicating that some fraction of the outflowing gas could be deposited into the galaxy halo. The outflows also exhibit large mass-loading factors (\blue{$\sim 1$--120}), and the estimated molecular gas depletion times, assuming a similar $\dot{M}_{\rm out}$ in the molecular phase, could explain recent observations of the sample's molecular gas content. Our results suggest that the observed outflows have the potential of quenching SF possibly by displacing the gas into the CGM and affecting the gaseous environment around the galaxies, consistent with recent studies of rapidly quenched galaxies at cosmic noon.
    
\end{enumerate}


The role of galactic outflows in both galaxy quenching and the maintenance of quiescence is complex and likely varies from case to case. Ultra-fast outflows capable of directly expelling gas from the host galaxies have been observed across cosmic time, and episodic AGN activity may also drive post-quenching outflows not necessarily related to the initial quenching event. Our study on a large statistical sample of PSBs at intermediate redshifts provides an important link between similar work in the local universe and at cosmic noon, highlighting the possibility that outflows may emerge as an important pathway for rapid quenching since $z \sim 0.5$. Future observations of gas in different phases will be essential for building a complete picture of galactic gas reservoirs, and spatially resolved studies of the CGM will further enlighten the ultimate fate of the outflowing material.

\begin{acknowledgments}
\blue{We thank the reviewer for the careful review and comments that improve the clarity of the manuscript.}

JSS, VRD, AG, and KAS gratefully acknowledge support from NSF-AAG\#2407954 and 2407955. VW acknowledges the Science and Technologies Facilities Council (ST/Y00275X/1) and Leverhulme Research Fellowship (RF-2024-589/4)

Funding for the Sloan Digital Sky Survey IV has been provided by the Alfred P. Sloan Foundation, the U.S. Department of Energy Office of Science, and the Participating Institutions. SDSS-IV acknowledges support and resources from the Center for High Performance Computing at the University of Utah. The SDSS website is www.sdss4.org. SDSS-IV is managed by the Astrophysical Research Consortium for the Participating Institutions of the SDSS Collaboration, including the Brazilian Participation Group, the Carnegie Institution for Science, Carnegie Mellon University, Center for Astrophysics | Harvard \& Smithsonian, the Chilean Participation Group, the French Participation Group, Instituto de Astrof\'isica de Canarias, The Johns Hopkins University, Kavli Institute for the Physics and Mathematics of the Universe (IPMU) / University of Tokyo, the Korean Participation Group, Lawrence Berkeley National Laboratory, Leibniz Institut f\"ur Astrophysik Potsdam (AIP), Max-Planck-Institut f\"ur Astronomie (MPIA Heidelberg), Max-Planck-Institut f\"ur Astrophysik (MPA Garching), Max-Planck-Institut f\"ur Extraterrestrische Physik (MPE), National Astronomical Observatories of China, New Mexico State University, New York University, University of Notre Dame, Observat\'ario Nacional / MCTI, The Ohio State University, Pennsylvania State University, Shanghai Astronomical Observatory, United Kingdom Participation Group, Universidad Nacional Aut\'onoma de M\'exico, University of Arizona, University of Colorado Boulder, University of Oxford, University of Portsmouth, University of Utah, University of Virginia, University of Washington, University of Wisconsin, Vanderbilt University, and Yale University.
\end{acknowledgments}

%
\facilities{Sloan}

\software{astropy \citep{2013A&A...558A..33A,2018AJ....156..123A,2022ApJ...935..167A},
pPXF \citep{Cappellari_2004,Cappellari_2017}, emcee \citep{mcmc}}


\appendix

\section{AGN vs. Non-AGN Stack} \label{appendix agn stack}

\begin{figure}[h]
\centering
\includegraphics[width=\columnwidth]{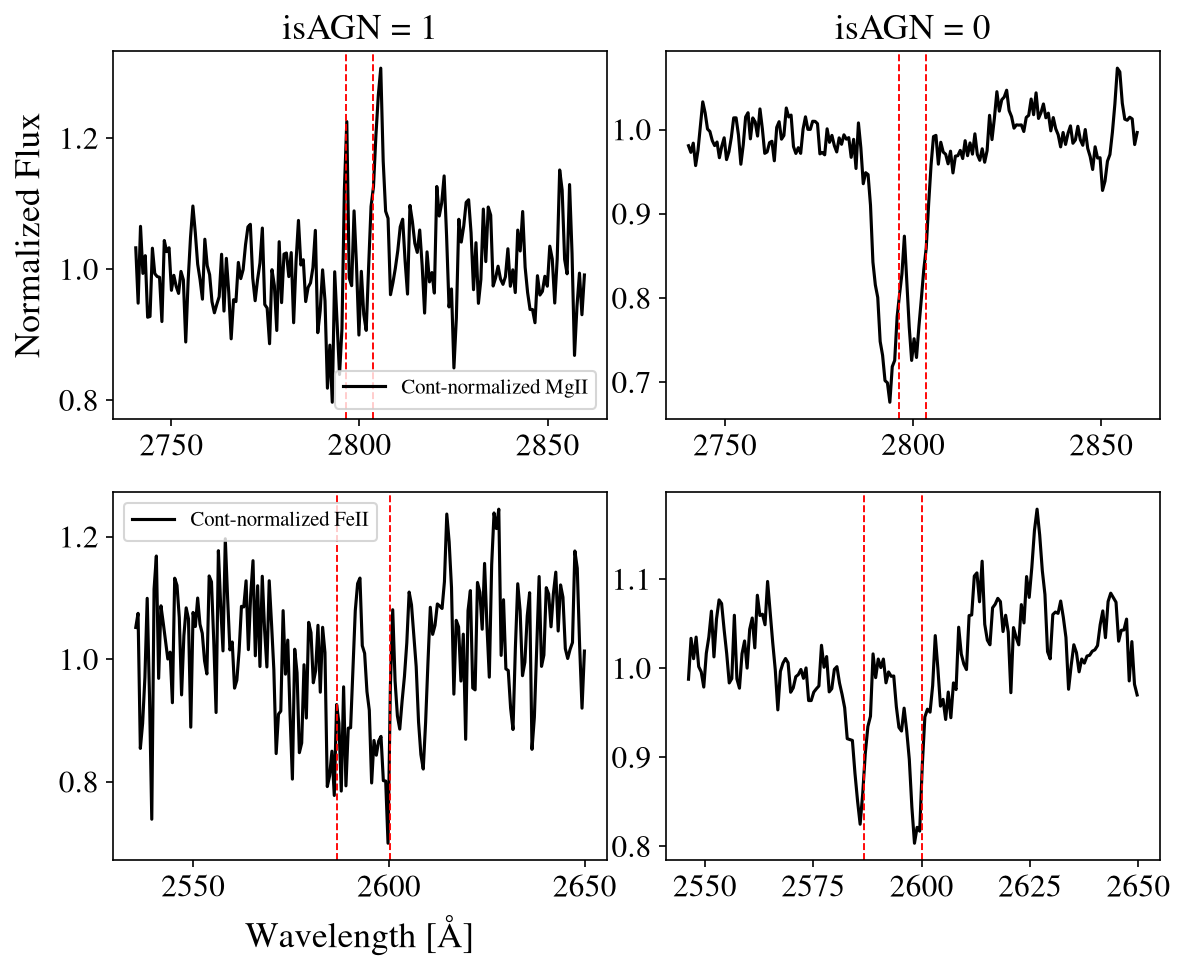}
\caption{Continuum normalized spectra for the AGN and no-AGN stack. The AGN stack exhibits little detectable ISM \feii\ and \mgii\ absorption.
\label{fig:agn_stack}}
\end{figure}

We present here (Figure \ref{fig:agn_stack}) the continuum normalized stacked spectra for galaxies that currently host/do not host AGN. See Section \ref{sec: outflow energetics} for more discussions.

\section{Outflow mass limit} \label{appendix mout}

\begin{figure*}
\centering
\includegraphics[width=\textwidth]{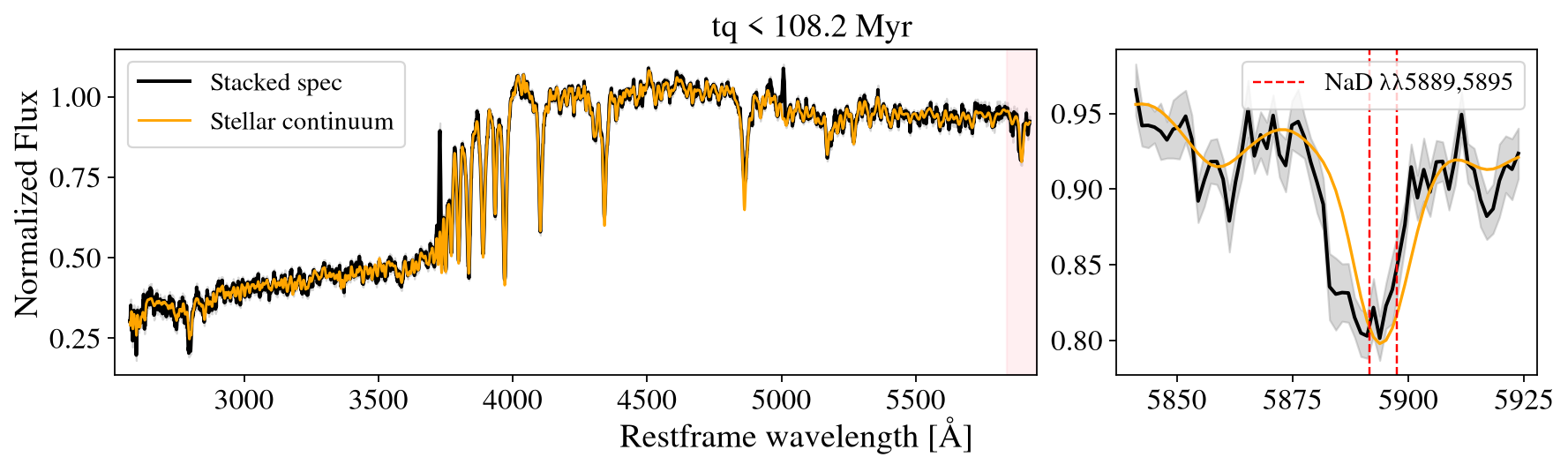}
\caption{\blue{Example of one stacked spectrum with \nad\ coverage (the smallest t$_q$ group) and corresponding stellar continuum. The left panel shows the entire wavelength range while the right panels show zoomed-in views of the \nad\ region (also shaded in pink in the left panel). The uncertainty of the stacked spectrum is plotted as gray shaded regions, and the restframe wavelengths of \nad\ lines are indicated by red vertical dashed lines. Unlike \feii\ and \mgii\ (Figure \ref{fig:ppxf_example}), \nad\ does not show significant gas absorption after removing the stellar continuum.} 
\label{fig:ppxf_example_nad}}
\end{figure*}

While we observe significant \feii\ and \mgii\ absorption from the ISM, we do not observe the same for \nad, which is another common neutral outflow tracer. \blue{An example of stacked spectra with \nad\ coverage is shown in Figure \ref{fig:ppxf_example_nad}.} Given the ionization potentials of neutral Mg, Fe, and Na (7.6, 7.9, and 5.1 eV, respectively) and those of \mgii\ and \feii\ (15.0 and 16.2 eV, respectively), \mgii\ and \feii\ can exist in neutral to weakly ionized gas, whereas \nad\ traces only neutral atomic or molecular gas. Consequently, \nad\ may probe a denser and cooler component of the outflow that contains mass too small to manifest in the spectra. Here we test this possibility by fitting mock spectra with manually added outflow components of varying mass and constrain the upper limit of mass in the \nad-traced outflows, if present.

\begin{figure}[h]
\centering
\includegraphics[width=\columnwidth]{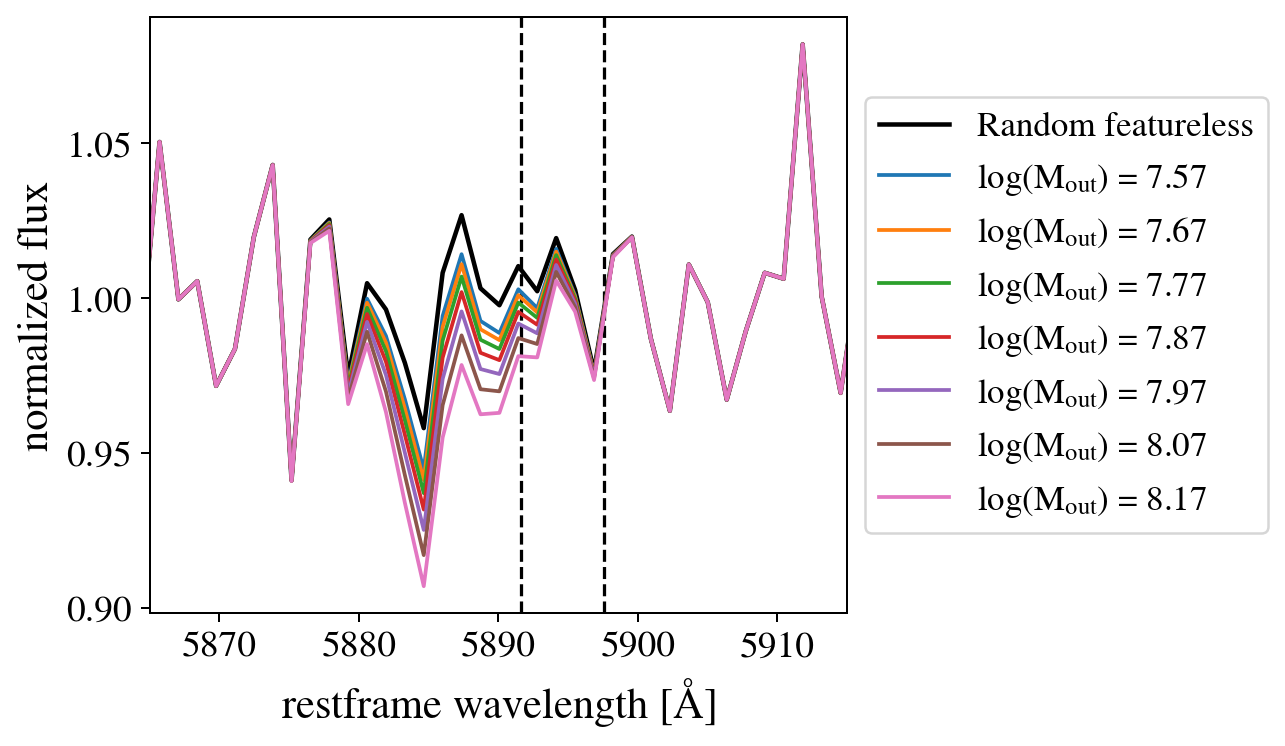}
\caption{Example random spectra for testing \nad-traced outflows with mock outflow signals in different outflow masses. \blue{The vertical dashed lines mark the restframe wavelengths of the \nad\ doublet.}
\label{fig:nad_test_spec}}
\end{figure}

\begin{figure}[h]
\centering
\includegraphics[width=\columnwidth]{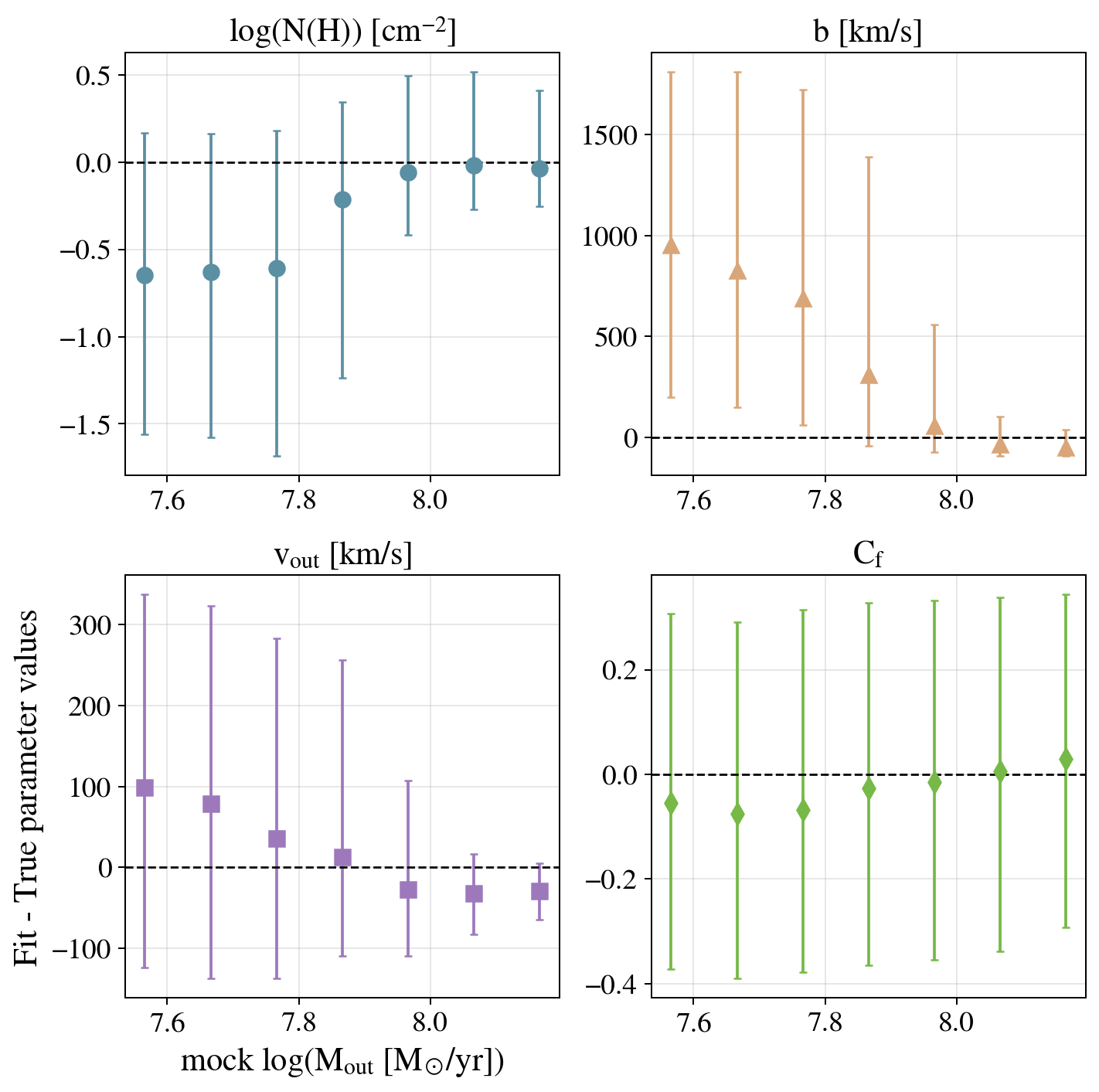}
\caption{The difference between the fitted and true values of the four free parameters in the one-component absorption model (Section \ref{sec: line model}) for the example spectra shown in Figure \ref{fig:nad_test_spec}. We repeat this process multiple times and find that the best agreement is achieved around log($M_{out} [M_{\odot}])\blue{\gtrsim}7.77-7.97$ in most cases. 
\label{fig:nad_test}}
\end{figure}

We generate featureless normalized spectra by randomly drawing the flux at each wavelength from a Gaussian distribution with a mean of 1 and a standard deviation of the flux uncertainty in a typical stacked spectra, and then multiplying these spectra by the one-component absorption profile defined in Section \ref{sec: line model} as the mock outflow signal (Figure \ref{fig:nad_test_spec}). In equation (3), we use $\lambda_0$ = 5897.558 \AA\ and $f$ = 0.3201, which are the rest-frame vacuum wavelength and oscillator strength of the $\lambda\lambda$5897 member of the \nad\ doublet \citep{Morton_2003}. We generate a set of absorption profiles spanning log(M$_{\rm out} [M_{\odot}])\sim7.57-8.17$, in 0.1 dex increment, by varying the column density parameter. In equation (7), we adopt the ionization fraction of 0.1, dust depletion of $-0.95$, and solar metallicity of $-5.69$ for Na \citep{Savage_1996}. The values of $v$, $C_f$, and $b_D$ are set to typical values of $-$350 km/s, 0.5, and 212 km/s, respectively. We then fit the mock spectra and examine how well the original parameters are recovered. Repeating this procedure multiple times, we find that in the majority of cases the true parameter values are recovered within the uncertainties, with the agreement between fitted and true parameters improving with increasing log($M_{out}$) and leveling off around log($M_{out} [M_{\odot}])\sim7.77-7.97$ (Figure \ref{fig:nad_test}). We therefore take this range as our estimated minimum outflow mass required for detection in our stacked spectra.

This $M_{out}$ limit from \nad\ is one order of magnitude smaller than that from \feii\ (Table \ref{tab:outflow_properties}). A similarly low $M_{out,Na}$ compared to those from Mg and Fe has also been observed in \citet{Valentino_2025}, where they attribute this difference to the \nad\ line being sensitive to gas conditions and preferentially tracing the colder and denser gas. It is also possible that the low ionization potential of Na requires a large amount of dust shielding to retain detectable amounts of neutral Na, and \squiggle\ PSBs may not be dusty enough. However, the stacking approach of this study complicates the interpretation and a detailed characterization of the dust content is beyond the scope of this work. The inferred limit of log($M_{out, Na} [M_{\odot}])\blue{\gtrsim}7.77-7.97$ is broadly consistent with the lowest value reported in \citet{Davies_2024}, log($M_{out, Na} [M_{\odot}])\sim7.01^{+0.43}_{-0.61}$, after accounting for the differences in conversion assumptions (\citealt{Davies_2024} adopted more conservative assumptions which produce masses $\sim0.54-$0.84 dex smaller than assumptions adopted here). It remains possible that \squiggle\ PSBs host \nad-traced neutral outflows similar to those observed in rapidly quenching galaxies at cosmic noon \citep[e.g.,][]{Davies_2024, Sun_2026}.

\section{Stacked spectra \blue{and modeling}} \label{appendix stack}
\begin{figure*}
\centering
\includegraphics[width=0.495\textwidth]
{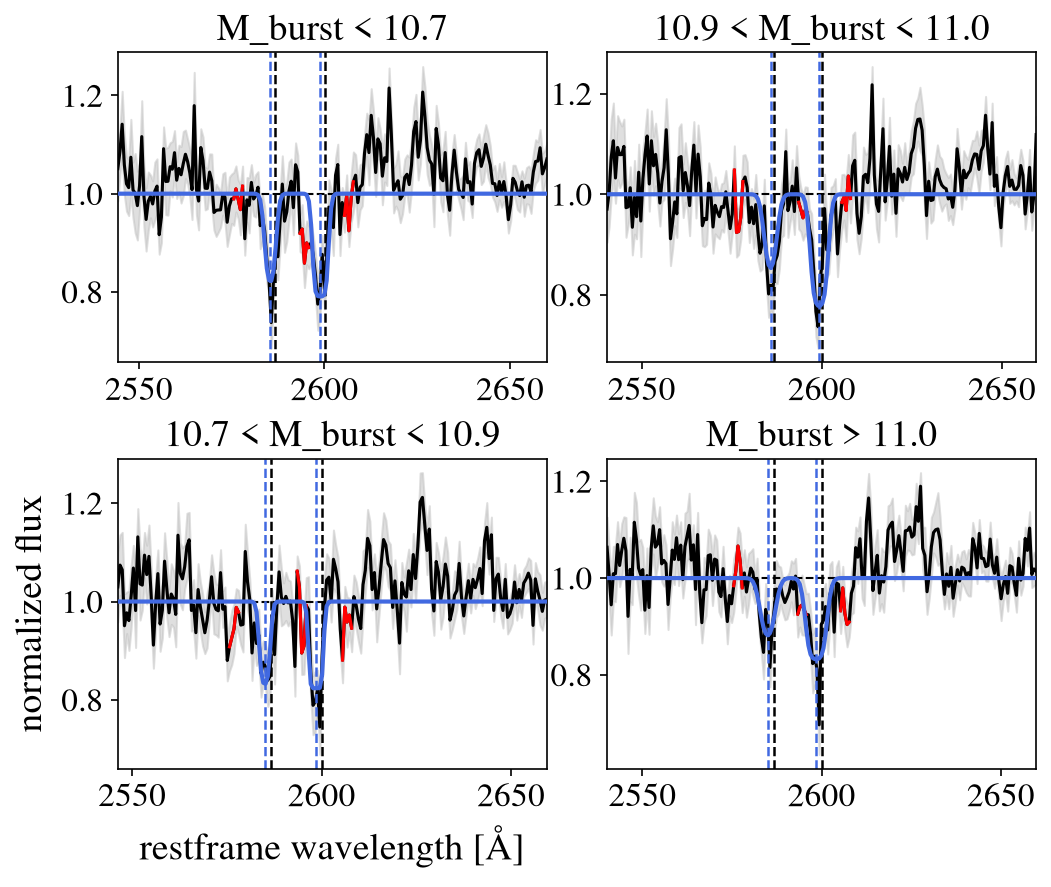}
\includegraphics[width=0.495\textwidth]
{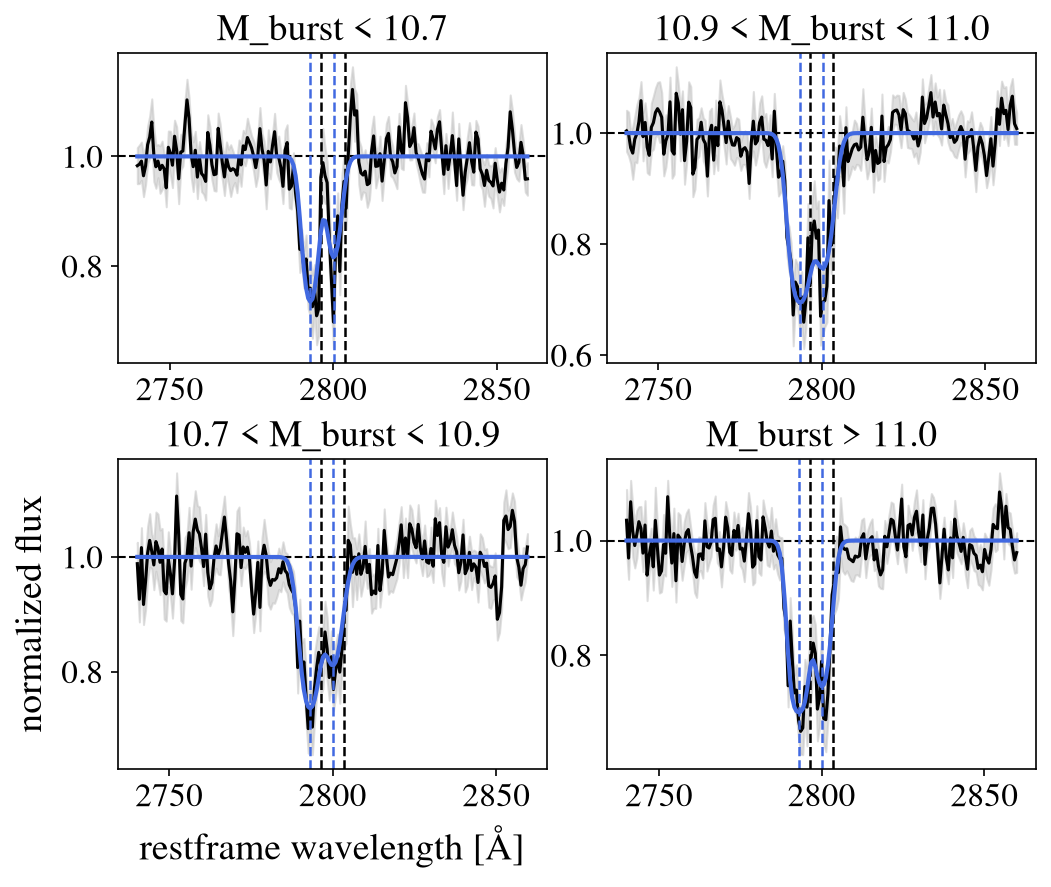}
\caption{\feii\ (left 4 panels) and \mgii\ (right 4 panels) profiles in the $M_{burst}$ stacks. The plotting scheme is the same as in Figure \ref{fig:tq_stack}.
\label{fig:Mburst_stack}}
\end{figure*}

\begin{figure*}
\centering
\includegraphics[width=0.495\textwidth]
{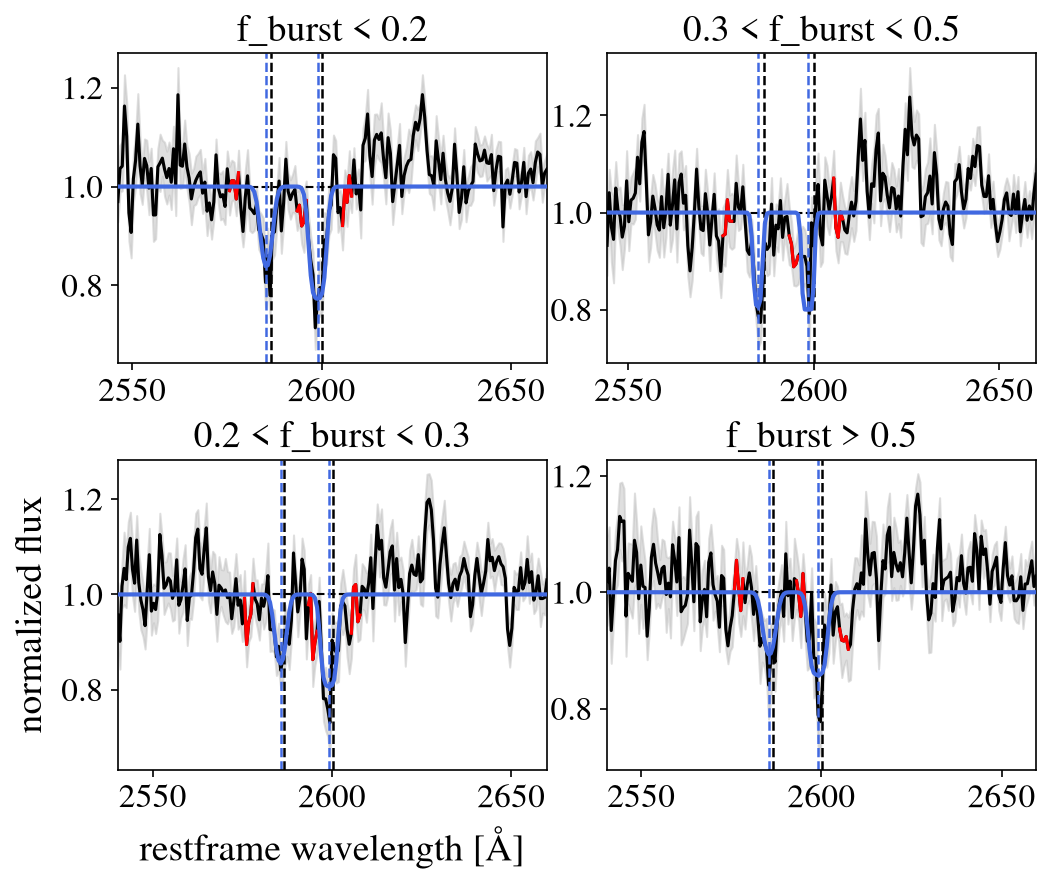}
\includegraphics[width=0.495\textwidth]
{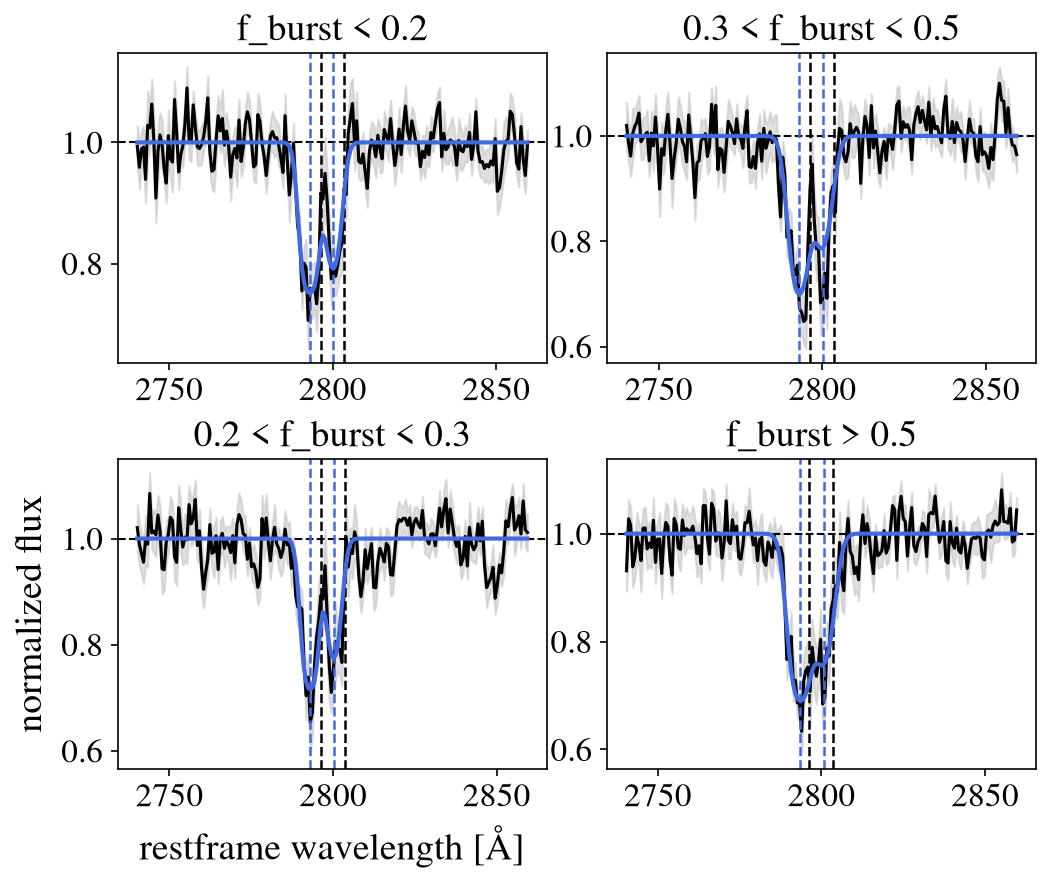}
\caption{\feii\ (left 4 panels) and \mgii\ (right 4 panels) profiles in the $f_{burst}$ stacks. The plotting scheme is the same as in Figure \ref{fig:tq_stack}.
\label{fig:fburst_stack}}
\end{figure*}

\begin{figure*}
\centering
\includegraphics[width=0.495\textwidth]
{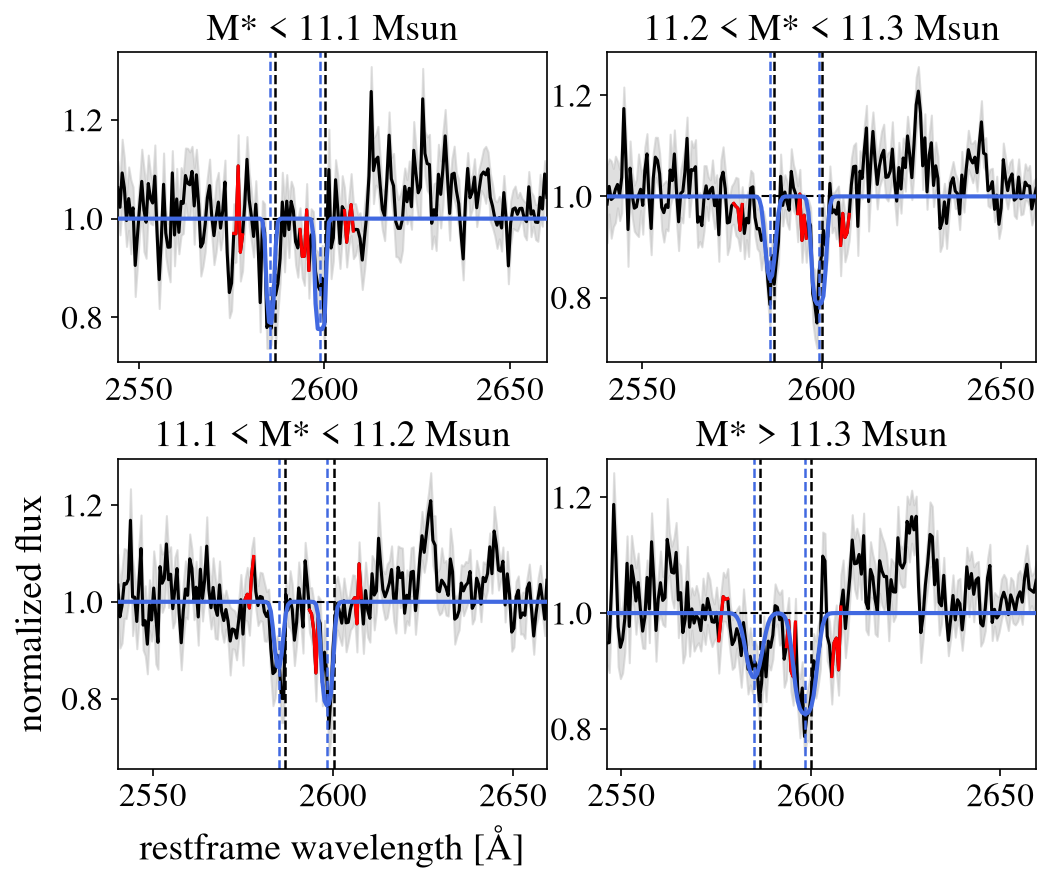}
\includegraphics[width=0.495\textwidth]
{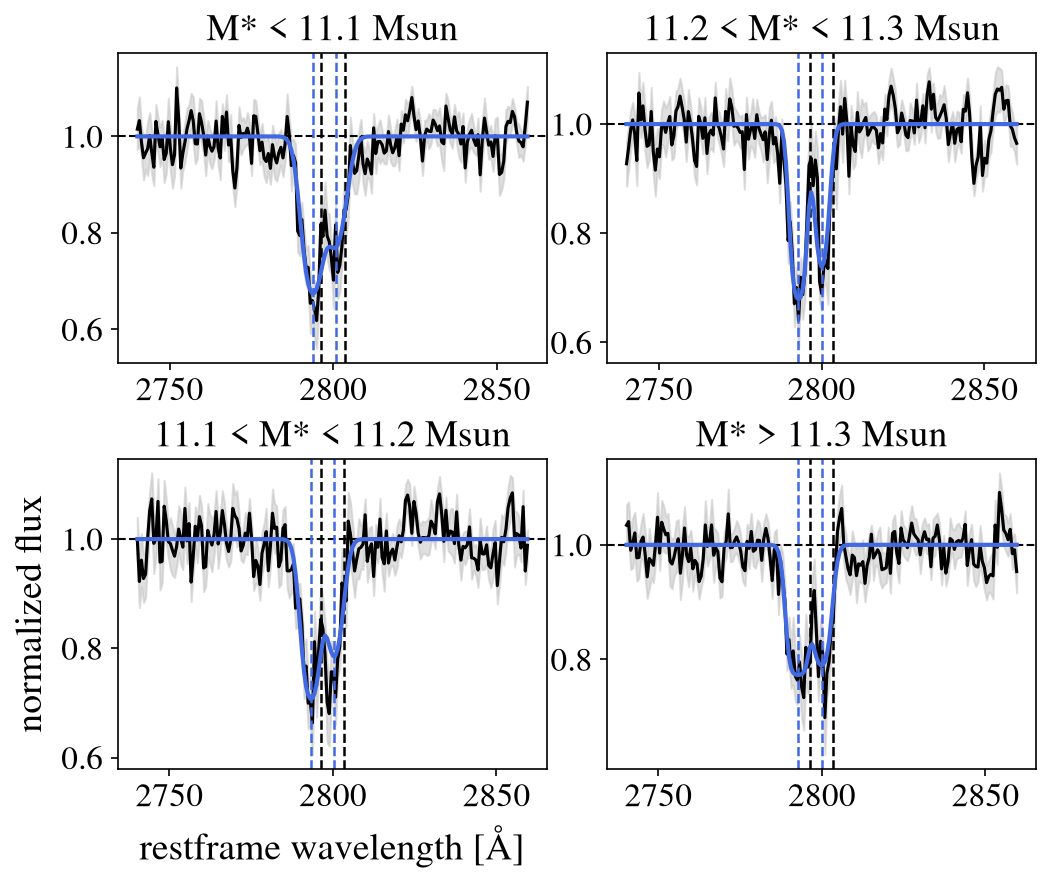}
\caption{\feii\ (left 4 panels) and \mgii\ (right 4 panels) profiles in the $M_*$ stacks. The plotting scheme is the same as in Figure \ref{fig:tq_stack}.
\label{fig:mass_stack}}
\end{figure*}

\begin{figure}
\centering
\includegraphics[width=0.498\columnwidth]
{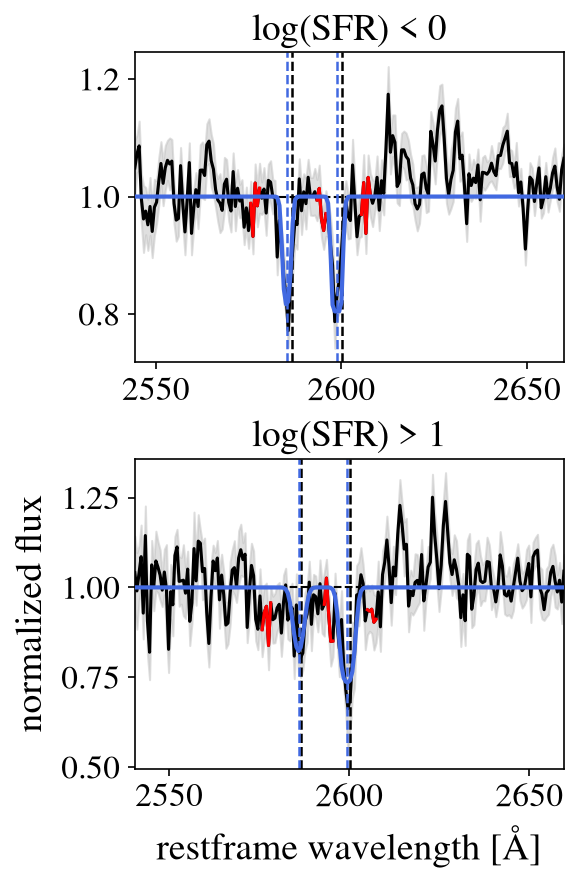}
\includegraphics[width=0.49\columnwidth]
{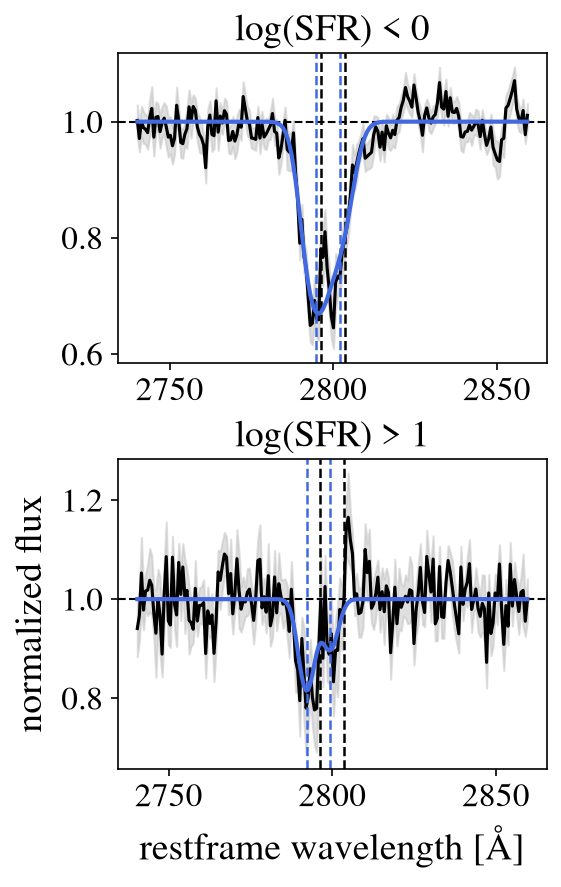}
\caption{\feii\ (left 2 panels) and \mgii\ (right 2 panels) profiles in the (s)SFR stacks. The plotting scheme is the same as in Figure \ref{fig:tq_stack}.
\label{fig:sfr_stack}}
\end{figure}

Here we present the stacked spectra overplotted with the best-fit one-component absorption model, zoomed in at the \feii\ and \mgii\ regions (Figure \ref{fig:Mburst_stack}-\ref{fig:sfr_stack}). 

\blue{Although we adopt wide prior ranges spanning the physically plausible values for all parameters in the line-profile models, we find a degeneracy between $N_{Fe}$ and $C_f$ in stacks where the \feii\ absorption is strongly saturated. Such saturated lines only weakly constrain $N_{Fe}$ and $C_f$ individually, since the intensity at line center is set primarily by $C_f$ rather than $N_{Fe}$. To break this degeneracy, we impose a physically motivated lower bound on the $C_f$ prior when the blue-to-red doublet ratio of \feii\ is $\sim1$. This $C_f$ lower bound is determined by $1-I_{min}$, where $I_{min}$ is the minimum observed normalized flux in the doublet after accounting for flux uncertainty. This bound follows from the partial-covering equation (Section \ref{sec: line model}) and ensures the fitted covering fraction is consistent with the depth of the observed saturated profile. This modification is not needed for \mgii, whose larger oscillator strength and smaller intrinsic blue-to-red doublet ratio (Section \ref{sec: outflow uncertainties}) allow both doublet members to reach the same saturated flux floor set by $C_f$ at comparatively modest column densities. \feii\ requires a much larger column density for both members to reach the same degree of saturation, and the fit may favor solutions with higher $N_{Fe}$ and lower $C_f$ that widen the absorption profile rather than deepen it if $C_f$ is left unconstrained.}

\blue{While model 1 is statistically preferred and adopted throughout this work for consistency, we find that the inferred outflow properties are generally robust to the choice of model in our cases. For \mgii, model 2, which includes an additional emission component, systematically yields slower outflow velocities and larger outflow masses, resulting in slightly lower mass outflow rates than model 1. The other models produce values that are generally scattered around and consistent with those from model 1. The small effect of a second absorption component is likely because the \mgii\ profiles are predominantly blueshifted relative to the systemic velocity. When an emission component is included, however, the cancellation between emission and absorption introduces additional complications that are difficult to disentangle. For \feii, model 3 yields outflow velocities that are generally consistent with, but occasionally higher than, those from model 1, while producing slightly lower outflow masses. This difference is not surprising given that the \feii\ profiles are less blueshifted than the \mgii\ profiles and therefore have a more prominent systemic component. Nevertheless, the two models yield comparable mass outflow rates for \feii-traced outflows.}

\bibliography{sample701,classics}{}
\bibliographystyle{aasjournalv7}


\end{CJK*}
\end{document}